%% file: main.tex
\documentclass{llncs}

\usepackage{xcolor}
\usepackage{amsmath,amssymb,amsfonts}
\usepackage{mathtools}
\usepackage{xspace}
\usepackage{wrapfig}
\usepackage{paralist}
\usepackage{microtype}
\usepackage{colonequals}
\usepackage{tikz}
\usepackage{cite}
\usepackage{multirow}
\usetikzlibrary{arrows.meta,decorations.pathmorphing,positioning,fit,trees,shapes,shadows,automata,calc,decorations.markings,patterns} 
\tikzset{>=latex}

\usepackage{algorithmicx,algorithm}
\usepackage[noend]{algpseudocode}
\usepackage{nicefrac}

\renewcommand{\emptyset}{\varnothing}

\usepackage{listings}

\include{prismstyle}
\input{macros}

\usepackage[caption=false,font=footnotesize]{subfig}

\begin{document}

\title{Counterexample-Driven Synthesis for \\ Probabilistic Program Sketches
\thanks{This work has been supported by the DFG RTG 2236 ``UnRAVeL'', the ERC Advanced Grant 787914 ``FRAPPANT''.}
}
\author{Milan \v{C}e\v{s}ka\inst{1}, Christian Hensel\inst{2},
Sebastian Junges\inst{2} and Joost-Pieter Katoen\inst{2}}
\institute{
Brno University of Technology, FIT, IT4I Centre of Excellence, Brno, Czech Republic
\and
RWTH Aachen University, Aachen, Germany 
}
\titlerunning{Revamping model repair}
\authorrunning{\v{C}e\v{s}ka, Dehnert, Jansen, Junges and Katoen}

\maketitle
\thispagestyle{plain}\pagestyle{plain}  

\input{00-abstract}

\section{Introduction}
\input{01-introduction}

\input{02-basics}

\section{CEGIS for Markov Chain Families}
\label{sec:CEGIS}
\input{03-CEGIS}

\input{03-approach}

\subsection{A program-level synthesiser}

\input{synthesiser}
\subsection{A program-level verifier}
\label{sec:highlevel:verifier}
\input{highlevelverifier}

\input{experiments}
\color{black}

\input{discussion}
\bibliographystyle{splncs04}
\bibliography{main,literature}

\clearpage
\appendix

\begin{center}
\textbf{\Large Appendix}
\end{center}

\section{Program-level Counterexamples for CEGIS}
\input{04-framework}


\end{document}

%% file: prismstyle.tex
\definecolor{prismgreen}{HTML}{009900}
\definecolor{prismred}{HTML}{cc0000}
\definecolor{prismblue}{HTML}{0000cc}

\lstdefinelanguage{Prism}{
        basicstyle=\color{prismred}\scriptsize\ttfamily,
        literate=*	{0}{{\textcolor{prismblue}{0}}}{1}
			{1}{{\textcolor{prismblue}{1}}}{1}
			{2}{{\textcolor{prismblue}{2}}}{1}
			{3}{{\textcolor{prismblue}{3}}}{1}
			{4}{{\textcolor{prismblue}{4}}}{1}
			{5}{{\textcolor{prismblue}{5}}}{1}
			{6}{{\textcolor{prismblue}{6}}}{1}
			{7}{{\textcolor{prismblue}{7}}}{1}
			{8}{{\textcolor{prismblue}{8}}}{1}
			{9}{{\textcolor{prismblue}{9}}}{1}
			{.0}{{\textcolor{prismblue}{.0}}}{2}
			{.1}{{\textcolor{prismblue}{.1}}}{2}
			{.2}{{\textcolor{prismblue}{.2}}}{2}
			{.3}{{\textcolor{prismblue}{.3}}}{2}
			{.4}{{\textcolor{prismblue}{.4}}}{2}
			{.5}{{\textcolor{prismblue}{.5}}}{2}
			{.6}{{\textcolor{prismblue}{.6}}}{2}
			{.7}{{\textcolor{prismblue}{.7}}}{2}
			{.8}{{\textcolor{prismblue}{.8}}}{2}
			{.9}{{\textcolor{prismblue}{.9}}}{2}
			{[}{{\textcolor{black}{[}}}{1}
			{]}{{\textcolor{black}{]}}}{1}
			{+}{{\textcolor{black}{+}}}{1}
			{-}{{\textcolor{black}{-}}}{1}
			{=}{{\textcolor{black}{=}}}{1}
			{<}{{\textcolor{black}{<}}}{1}
			{>}{{\textcolor{black}{>}}}{1}
			{\&}{{\textcolor{black}{\&}}}{1}
			{|}{{\textcolor{black}{|}}}{1}
			{:}{{\textcolor{black}{:}}}{1}
			{;}{{\textcolor{black}{;}}}{1}
			{(}{{\textcolor{black}{(}}}{1}
			{)}{{\textcolor{black}{)}}}{1}
			{..}{{\textcolor{black}{..}}}{2},
        keywords= {bool,ceil,const,ctmc,double,dtmc,endinit,endmodule,endrewards, endsystem,F,false,floor,formula,G,global,I,init,int,label,max,mdp,min,module,nondeterministic,P,Pmin,Pmax,prob,probabilistic,rate,rewards,Rmin,Rmax,S,stochastic,system,true,U, option, either, assignment, relation, operation, hole, variable},
        keywordstyle={\bfseries\color{black}},
        numberstyle=\footnotesize\color{black},
        comment=[l] {//}, morecomment=[s]{/*}{*/},
        commentstyle= \color{prismgreen},
        tabsize=4,
        captionpos=b,
        escapechar=^,
        moredelim=[is][\color{orange}]{@}{@},
}

%% file: macros.tex
\newcommand{\init}{\ensuremath{s_0}}

\newcommand{\Succ}{\ensuremath{\mathsf{Succ}}}

\usepackage{algorithmicx,algorithm}
\usepackage[noend]{algpseudocode}

\newcommand{\familymc}{\ensuremath{\mathfrak{D}}}
\newcommand{\semantics}[1]{[\![\, #1 \, ]\!]}

\newcommand{\act}{\ensuremath{a}}

\newcommand{\Distr}{\mathit{Distr}}

\newcommand{\distDom}{X}
\newcommand{\distFunc}{\mu}
\newcommand{\distDomElem}{x}

\newcommand{\dtmc}{\ensuremath{D}}

\newcommand{\probdtmc}{\ensuremath{P}}
\newcommand{\probfam}{\ensuremath{\mathfrak{P}}}

\algnewcommand{\IIf}[1]{\State\algorithmicif\ #1\ \algorithmicthen}
\algnewcommand{\EElse}[1]{\State\algorithmicelse\ #1\ }
\algnewcommand{\EndIIf}{\unskip}
\algrenewcommand\algorithmicindent{0.8em}%

\def\BibTeX{{\rm B\kern-.05em{\sc i\kern-.025em b}\kern-.08em
    T\kern-.1667em\lower.7ex\hbox{E}\kern-.125emX}}

\newcommand{\squishlist}{
   \begin{list}{$\bullet$}
    { \setlength{\itemsep}{2pt}    \setlength{\parsep}{0pt}
      \setlength{\topsep}{3pt}     \setlength{\partopsep}{0pt}
      \setlength{\leftmargin}{1.35em} \setlength{\labelwidth}{1em}
      \setlength{\labelsep}{0.5em} } }

\newcommand{\squishend}{
    \end{list}  }

\newcommand{\Option}{\mathsf{Option}}
\newcommand{\Sketch}{\mathfrak{S}_H}

\newcommand{\program}[2]{\mathcal{#1}_{#2}}
\newcommand{\SketchConstraints}{\ensuremath{\Gamma}}
\newcommand{\OptionPrices}{\ensuremath{\mathsf{cost}}}

\newcommand{\Vars}{\mathrm{Var}}

\newcommand{\Prob}{\mathsf{Prob}}

\newcommand{\submc}[2]{\ensuremath{{#1}\!\downarrow\!{#2}}}

\newcommand{\fixdeadlock}{\mathsf{fixdl}}

\newcommand{\queue}{\ensuremath{Q}}

%% file: 00-abstract.tex
\begin{abstract}
Probabilistic programs are key to deal with uncertainty in e.g.\ controller synthesis. They are typically small but intricate. Their development is complex and error prone requiring quantitative reasoning over a myriad of alternative designs. To mitigate this complexity, we adopt counterexample-guided inductive synthesis (CEGIS) to automatically synthesise finite-state probabilistic programs. Our approach leverages efficient model checking, modern SMT solving, and counterexample generation at program level. Experiments on practically relevant case studies show that design spaces with millions of candidate designs can be fully explored using a few thousand verification queries.
\end{abstract}

%% file: 01-introduction.tex
With the ever tighter integration of computing systems with their environment, quantifying (and minimising) the probability of encountering an anomaly or unexpected behaviour becomes crucial. 
This insight has led to a growing interest in probabilistic programs and models in the software engineering community.
Henzinger~\cite{DBLP:journals/ife/Henzinger13} for instance argues that ``the Boolean partition of software into correct and incorrect programs falls short of the practical need to assess the behaviour of software in a more nuanced fashion $[\hdots]$.''
In \cite{DBLP:conf/kbse/Rosenblum16}, Rosenblum advocates taking a more probabilistic approach in software engineering.
Concrete examples include quantitative analysis of software product lines~\cite{DBLP:journals/infsof/GhezziS13,DBLP:conf/splc/VarshosazK13,RodriguesANLCSS15,DBLP:journals/fac/ChrszonDKB18,DBLP:conf/fm/VandinBLL18}, synthesis of probabilities for adaptive software~\cite{JSS17,7177126}, and probabilistic model checking at runtime to support verifying dynamic reconfigurations~\cite{DBLP:journals/cacm/CalinescuGKM12,DBLP:journals/tse/FilieriTG16}.

\noindent\emph{Synthesis of probabilistic programs.}
The development of systems under uncertainty is intricate. Probabilistic programs are a prominent formalism to deal with uncertainty.
Unfortunately, such programs are rather intricate.
Their development is complex and error prone requiring quantitative reasoning over many alternative designs.
One remedy is the exploitation of probabilistic model checking~\cite{DBLP:reference/mc/BaierAFK18} using 
a \emph{Markov chain} as the operational model of a program.
One may then apply model checking on each design, or some suitable representation thereof~\cite{DBLP:journals/fac/ChrszonDKB18,DBLP:conf/tacas/CeskaJJK19}.
Techniques such as parameter synthesis~\cite{hahn2011probabilistic,Ceska2017,DBLP:conf/atva/QuatmannD0JK16} and model repair~\cite{DBLP:conf/tacas/BartocciGKRS11,DBLP:conf/tase/ChenHHKQ013} have been successful, but they only allow to amend or infer transition probabilities, whereas the control structure---the topology of the probabilistic model---is fixed.

\begin{wrapfigure}[8]{r}{51mm}
\centering
  \tikzset{
        hatch distance/.store in=\hatchdistance,
        hatch distance=6pt,
        hatch thickness/.store in=\hatchthickness,
        hatch thickness=1pt
    }
    \makeatletter
    \pgfdeclarepatternformonly[\hatchdistance,\hatchthickness]{flexible hatch}
    {\pgfqpoint{0pt}{0pt}}
    {\pgfqpoint{\hatchdistance}{\hatchdistance}}
    {\pgfpoint{\hatchdistance-1pt}{\hatchdistance-1pt}}%
    {
        \pgfsetcolor{\tikz@pattern@color}
        \pgfsetlinewidth{\hatchthickness}
        \pgfpathmoveto{\pgfqpoint{0pt}{0pt}}
        \pgfpathlineto{\pgfqpoint{\hatchdistance}{\hatchdistance}}
        \pgfusepath{stroke}
    }
    \makeatother
    \vspace{-2em}
    \scalebox{0.7}{
      \begin{tikzpicture}
      \tikzset{inv/.style={inner sep=0pt, minimum size=0pt}}
      \newcommand{\tikzEdgeLength}{1.2cm}
      \newcommand{\tikzsmallerfontsize}{\footnotesize}
    
      \node[minimum width=\tikzEdgeLength, minimum height=\tikzEdgeLength, draw] (synthesiser) {\small {Synth\phantom{esiser}}};
          
      \node[right=2*\tikzEdgeLength of synthesiser.east, minimum width=\tikzEdgeLength, minimum height=\tikzEdgeLength, draw, minimum width=\tikzEdgeLength, rounded corners] (verifier) {\small Verifier};
      
      \path[pattern color=lightgray,pattern=flexible hatch] ($(synthesiser.north east)-(0.01,0.01)$) -- ($(synthesiser.north east)-(0.4,0.01)$) to[draw,  bend right] ($(synthesiser.south)+(0.15,0.01)$) -- ($(synthesiser.south east)+(-0.01,0.01)$) -- cycle;
      \draw ($(synthesiser.north east)-(0.4,0)$) to[draw,  bend right] ($(synthesiser.south)+(0.15,0.01)$);
      
      \node[inv, circle, inner sep=0.7pt, fill=black] (instance) at ($(synthesiser.north east)-(0.3,0.5)$) {};
      
      \draw[->, shorten <= 1pt, shorten >= 1pt] (instance) (instance) to[bend left] node[auto, pos=0.6, yshift=-1pt] {\tikzsmallerfontsize instance} (verifier);
      \draw[->] (verifier) to[bend left] node[auto, swap, near start, xshift=5pt, yshift=-3pt] {\tikzsmallerfontsize reject +} node[auto, xshift=10pt, yshift=2pt] (cex) {\tikzsmallerfontsize CE} ($(instance)-(0.2,0.4)$);
      
      \coordinate (topleft) at ($(synthesiser.north west)+(-0.3,0.4)$);
      \coordinate (bottomright) at ($(verifier.south east)+(0.3,-0.55)$);

      \draw[densely dotted, rounded corners] (topleft) rectangle (bottomright);
      
      \node[above=0.55cm of synthesiser] (sketch) {\tikzsmallerfontsize sketch};
      \draw[->] (sketch) to (synthesiser);

      \node (properties) at (verifier |- sketch) {\tikzsmallerfontsize properties};
      \draw[->] (properties) to (verifier);

      \node[below=0.75cm of synthesiser] (unsat) {\tikzsmallerfontsize unsatisfiable};
      \draw[->] (synthesiser) to node[near start, xshift=9pt] {\tikzsmallerfontsize no instance} (unsat);

      \node[align=center, anchor=north] (synthprog) at (verifier |- unsat.north) {\tikzsmallerfontsize synthesised program};
      \draw[->] (verifier) to node[left] {\tikzsmallerfontsize accept} (synthprog);
      
    \end{tikzpicture}
    }
    \vspace{-1em}
  \caption{CEGIS for synthesis.}
  \label{fig:CEGIS}

\end{wrapfigure}
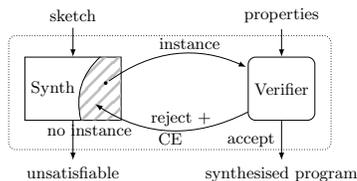

\smallskip\noindent\emph{Counter-Example-Guided Inductive Synthesis}.
This paper aims to overcome the existing limitation, by adopting the paradigm of \emph{CounterExample-Guided Inductive Synthesis} (CEGIS, cf.~Fig.~\ref{fig:CEGIS})~\cite{Solar-LezamaASPLOS2006,sygus,DBLP:conf/pldi/Solar-LezamaRBE05,DBLP:conf/cav/AbateDKKP18} to finite-state probabilistic models and programs.
Program synthesis amounts to automatically provide an instantiated probabilistic program satisfying all properties, or returns that such realisation is non-existing.
This syntax-based approach starts with a sketch, a program with holes, and iteratively searches for good---or even optimal---realisations, i.e., instantiated programs.
Rather than checking all realisations, the design space is pruned by potentially ruling out many realisations (dashed area) at once.
From every realisation that was verified and rejected, a counterexample (CE) is derived, e.g., a program run violating the specification. 
An SMT (satisfiability modulo theory)-based synthesiser uses the CE to prune programs that also violate the specification.  
These programs are safely removed from the design space. 
The synthesis and verification step are repeated until either a satisfying program is found or the entire design space is pruned implying the non-existence of such a program.

\smallskip
\noindent\emph{Problem statement and program-level approach.}
This paper tailors and generalises CEGIS to probabilistic models and programs.
The input is a sketch---a probabilistic program with holes, where each hole can be replaced by finitely many options---, a set of quantitative properties that the program needs to fulfil, and a  budget.
All possible realisations have a certain cost and the synthesis  provides a realisation that fits within the budget.
Programs are represented in the PRISM modelling language~\cite{KNP11} and properties are expressed in PCTL (Probabilistic Computational Tree Logic) extended with rewards, 
as standard in probabilistic model checking~\cite{KNP11,DBLP:conf/cav/DehnertJK017}. 
Program sketches succinctly describe the design space of the system by providing the program-level structure but leaving some parts (e.g., command guards or variable assignments) unspecified. 


\smallskip
\noindent\emph{Outcomes.}
To summarise, this paper presents a novel synthesis framework for probabilistic programs that adhere to a given set of quantitative requirements and a given budget. 
We use families of Markov chains to formalise our problem, 
and then formulate a CEGIS-style algorithm on these families. 
Here, CEs are subgraphs of the Markov chains. 
In the second part, we then generalise the approach to reason on probabilistic programs with holes.
While similar in spirit, we rely on program-level CEs~\cite{DBLP:conf/atva/DehnertJWAK14,DBLP:journals/corr/abs-1305-5055}, and allow for a more flexible sketching language.
To the best of our knowledge, this is the first lifting of CEGIS to probabilistic programs.
The CEGIS approach is sound and complete: either an admissible program does exist and it is computed, or no such program exists and the algorithm reports this.
We provide a prototype implementation build on top of the model checker Storm~\cite{DBLP:conf/cav/DehnertJK017} and the SMT-tool Z3~\cite{Z3}.
Experiments with different examples demonstrate scalability: design spaces with millions of realisations can be fully explored by a few thousand verification queries and result in a speedup of orders of magnitude.

\medskip
\noindent\textbf{Related work.}
We build on the significant body of research that employs formal methods to analyse quality attributes of alternative designs, e.g.\ \cite{balsamo2004model,bondy2014,Becker20093,performability2016,stewart2009probability,woodside-etal2014}. Enumerative approaches based on 
 Petri nets \cite{lindemann1998performance}, stochastic models \cite{7177126,Sharma2007493} and timed automata \cite{hessel2008testing,larsen2014verification}, and the corresponding tools  for simulation and verification (e.g.\ Palladio~\cite{Becker20093}, PRISM~\cite{KNP11}, UPPAAL~\cite{hessel2008testing}) have long been used.
 
For non-probabilistic systems, CEGIS can find  programs for a variety of challenging problems~\cite{DBLP:conf/pldi/Solar-LezamaRBE05,Solar-LezamaPLDI2008}.
Meta-sketches and the \emph{optimal and quantitative synthesis problem} in a non-probabilistic setting have been proposed~\cite{vcerny2011quantitative,QuantSyn,BornholtPOPL2016}.

A prominent representation of sets of alternative designs are modal transition systems~\cite{DBLP:conf/lics/LarsenT88,DBLP:journals/eatcs/AntonikHLNW08,DBLP:conf/birthday/Kretinsky17}. 
In particular, \emph{parametric} modal transition systems~\cite{DBLP:journals/acta/BenesKLMSS15} and synthesis therein~\cite{DBLP:conf/lpar/BenesKLMS12} allow for similar dependencies that occur in program-level sketches. 
Probabilistic extensions are considered in, e.g.~\cite{DBLP:journals/iandc/DelahayeKLLPSW13}, but not in conjunction with synthesis. 
Recently~\cite{DBLP:conf/tacas/DurejaR18} 
proposed to exploit relationships between model and specification, thereby reducing  the  number of model-checking instances. 

%
In the domain of quantitative reasoning, sketches and likelihood computation are used  to find probabilistic programs that best match available data~\cite{NoriPLDI2015}. 
The work closest to our approach synthesises  probabilistic systems from specifications and parametric templates~\cite{DBLP:conf/kbse/GerasimouTC15}. 
The principal difference to our approach is the use of counterexamples.
The authors leverage evolutionary optimisation techniques  without pruning. Therefore, the completeness is only achieved by exploring all designs, which is practically infeasible.
%
An extension to handle parameters  affecting transition probabilities (rates) has been integrated into the evolutionary-driven synthesis~\cite{ICSA17,JSS17} and is available in RODES~\cite{QEST17}.
Some papers have considered the analysis of sets of alternative designs within the quantitative verification of software product lines~\cite{DBLP:journals/infsof/GhezziS13,DBLP:conf/splc/VarshosazK13,RodriguesANLCSS15}.
The typical approach is to  analyse all individual designs (product configurations) or build and analyse a single (so-called \emph{all-in-one}) Markov decision process describing all the designs simultaneously. Even with symbolic methods, this hardly scales to large sets of alternative designs.
These techniques have recently been integrated into ProFeat~\cite{DBLP:journals/fac/ChrszonDKB18} and QFLan~\cite{DBLP:conf/fm/VandinBLL18}.
An abstraction-refinement scheme has recently been explored in~\cite{DBLP:conf/tacas/CeskaJJK19}. It iteratively analyses an abstraction of a (sub)set of designs---it is an orthogonal and slightly restricted approach to the inductive method presented here (detailed differences are discussed later).
An incomplete method in~\cite{DBLP:journals/corr/abs-1807-06106} employs abstraction targeting a particular case study.
SMT-based encodings for synthesis in Markov models have been used in, e.g.~\cite{DBLP:conf/tacas/Junges0DTK16,Cardelli2017}. 
These encodings are typically monolithic---they do not prune the search space via CEs.
Probabilistic CEs have been recently used to ensure that controllers obtained via 
learning from positive examples meet given safety properties~\cite{Wenchao2018}. 
In contrast, we leverage program-level CEs that can be 
used to prune the design space.


%% file: 02-basics.tex
\section{Preliminaries and Problem Statement}

We start with basics of probabilistic model checking, for details, see~\cite{BK08,DBLP:reference/mc/BaierAFK18}, and then formalise  families of Markov chains. 
Finally, we define some synthesis problems.

\medskip\noindent\textbf{Probabilistic models and specifications.}
A \emph{probability distribution} over a finite set $\distDom$
is a function $\distFunc\colon\distDom\rightarrow [0,1]$ with $\sum_{\distDomElem\in\distDom}\distFunc(\distDom)=1$.
Let $\Distr(\distDom)$ denote the set of all distributions on $\distDom$. 
\begin{definition}[MC]\label{def:dtmc}
A \emph{discrete-time Markov chain} (MC) $\dtmc$ is a tuple $(S,\init, \probdtmc)$ with finite set $S$ of states, initial state $\init \in S$, and transition probabilities $\probdtmc\colon S \rightarrow \Distr(S)$. We write $P(s,t)$ to denote $P(s)(t)$.
\end{definition}
For $S' \subseteq S$, the set ${\Succ(S') \colonequals \{ t \in S \mid \exists s \in S'.~\probdtmc(s,t) > 0\}}$ denotes the successor states of $S'$.
A \emph{path} of an MC $D$ is an (in)finite sequence $\pi = s_0s_1s_2$,
where $s_i\in S$, and $s_{i+1} \in \Succ(s_i)$ 
for all $i\in\mathbb{N}$.
\begin{definition}[sub-MC]
Let MC $D = (S,\init, \probdtmc)$ and  \emph{critical states} $C \subseteq S$ with $\init \in C$.
The \emph{sub-MC} of $D,C$ is the MC $\submc{D}{C} = (C \cup \Succ(C), \init, \probdtmc')$ with $P'(s,t) = P(s,t)$ for $s \in C$, $P'(s,s) = 1$ for $s \in \Succ(C) {\setminus} C$, and $P'(s,t) = 0$

\end{definition}

\smallskip\noindent\emph{Specifications.} 
For simplicity, we focus on reachability properties ${\varphi=\mathbb{P}_{\sim \lambda}(\lozenge G)}$ for a set $G \subseteq S$ of goal states, threshold $\lambda\in [0,1]\subseteq{\mathbb{R}}$, and comparison relation ${{}\sim{}} \in \{<,\leq,\geq,>\}$. 
The interpretation of $\varphi$ on MC $\dtmc$ is as follows.
Let $\texttt{Prob}(\dtmc,\lozenge G)$ denote the probability to reach $G$ from $\dtmc$'s initial state.
Then, $\dtmc\models\varphi$ if $\texttt{Prob}(\dtmc, \lozenge G) \sim \lambda$.
A specification is a set $\Phi = \{\varphi_i\}_{i\in I}$ of properties, and $D \models \Phi$ if $\forall i \in I.~D \models \varphi_i$.
Upper-bounded properties (with ${{}\sim{}}\in \{ <, \leq \}$) are safety properties, lower-bounded properties are liveness properties.
Extensions to expected rewards  
or $\omega$-regular properties are rather straightforward. 

\medskip\noindent\textbf{Families of Markov chains.}
We recap an explicit representation of a \emph{family of MCs} using a parametric transition function, as in~\cite{DBLP:conf/tacas/CeskaJJK19}.
\begin{definition}[Family of MCs]
A \emph{family of MCs} is a tuple $\familymc =(S,\init,K,\mathfrak{P})$ with $S$, $\init$ as before, a finite set  of parameters $K$ where the domain for each parameter $k\in K$ is $T_k\subseteq S$, and transition probability function $\mathfrak{P}\colon S \rightarrow \Distr(K)$. 
\end{definition}
The transition probability function of MCs maps states to distributions over successor states. 
For families, this function maps states to distributions over parameters.
Instantiating each parameter with a value from its domain yields a ``concrete'' MC, called a \emph{realisation}.

\begin{definition}[Realisation]
	A \emph{realisation} of a family $\familymc=(S,\init,K,\mathfrak{P})$ is a function $r\colon K \rightarrow S$ where $\forall k\in K\colon r(k) \in T_k$. 
	A realisation $r$ yields an MC $D_r \colonequals (S,\init,\mathfrak{P}(r))$, where $\mathfrak{P}(r)$ is the transition probability matrix in which 
	each  
	$k\in K$ in $\mathfrak{P}$ 
	is replaced by $r(k)$.
	Let $\mathcal{R}^\familymc$ denote the \emph{set of all realisations} for $\familymc$.
\end{definition}
As a family $\familymc$ has finite parameter domains, the number of family members (i.e.\ realisations from $\mathcal{R}^\familymc$) of $\familymc$ is finite, 
but exponential in $|K|$.
While all MCs share their state space, their \emph{reachable} states may differ. 
\begin{example}\label{ex:dtmc_family}	Consider the family of MCs $\familymc = (S, \init, K, \mathfrak{P})$
	where $S=\{0,\hdots,4\}$, $\init = 0$, and $K=\{k_0, \hdots, k_5\}$ with  
	$T_{k_0} = \{ 0 \}$, $T_{k_1} = \{ 1 \}$, $T_{k_2} = \{ 2, 3 \}$, $T_{k_3} = \{ 2, 4 \}$, $T_{k_4} = \{ 3 \}$ and $T_{k_5} = \{ 4 \}$, and $\mathfrak{P}$ given by:
	\begin{align*}
	&	\mathfrak{P}(0) = 0.5 \colon k_1 + 0.5 \colon k_2 & 
	& \mathfrak{P}(1) = 0.1 \colon k_0 + 0.8 \colon k_3 + 0.1 \colon k_5 & 
   &  \mathfrak{P}(2) = 1\colon k_3 \\ 
 & \mathfrak{P}(3) = 1 \colon k_4 & 
&	\mathfrak{P}(4) = 1\colon k_5 &	
	\end{align*}
	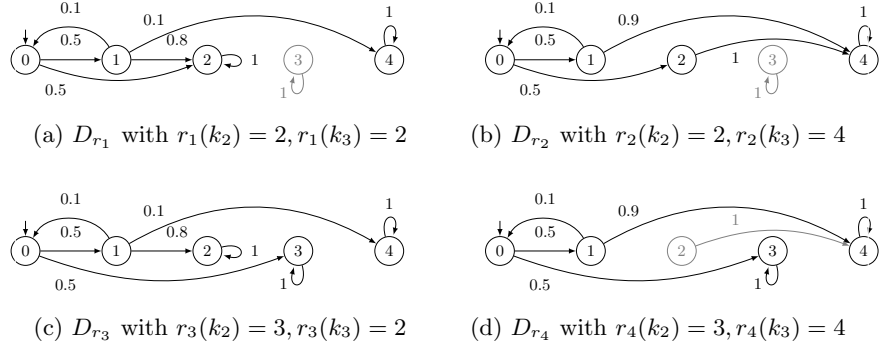
\begin{figure}[t]
	\centering
	\subfloat[$D_{r_1}$ with $r_1(k_2)=2, r_1(k_3)=2$]{%
	    \label{fig:realisation_0_2}
		\scalebox{0.68}{\input{pics/realisation_2_2}}}
	\subfloat[$D_{r_2}$ with $r_2(k_2)=2, r_2(k_3)=4$]{
	    \label{fig:realisation_1_2}
		\scalebox{0.68}{\input{pics/realisation_2_4}}} \\
	\subfloat[$D_{r_3}$ with $r_3(k_2)=3, r_3(k_3)=2$]{%
	    \label{fig:realisation_1_3}
		\scalebox{0.68}{\input{pics/realisation_3_2}}}
	\subfloat[$D_{r_4}$ with $r_4(k_2)=3, r_4(k_3)=4$]{%
	    \label{fig:realisation_0_3}
		\scalebox{0.68}{\input{pics/realisation_3_4}}}
		\vspace{-0.5em}
		\caption{The four different realisations of family $\familymc$.}
		
		\label{fig:realisation}
\end{figure}
Fig.~\ref{fig:realisation} shows the four MCs
of $\familymc$.
Unreachable states are greyed out.
\end{example}
The function $c\colon \mathcal{R}^{\familymc} \to \mathbb{N}$ assigns \emph{realisation costs}. Attaching costs to realisations is a natural way to distinguish preferable realisations.
We stress the difference with rewards in MCs; the latter impose a cost structure on paths in MCs.

\medskip\noindent\textbf{Problem statement}
\label{sec:problemstatement}
\emph{Synthesis problems.} Let $\familymc$ be a family, and $\Phi$ be a set of properties, and $B \in \mathbb{N}$ a budget. Consider the synthesis problems:
\begin{enumerate}
\item \emph{Feasibility synthesis:}
Find a realisation $r \in \mathcal{R}^{\familymc}$ with $D_r \models \Phi$ and $c(r) \leq B$.
\item \emph{Max synthesis:}
For given $G \subseteq S$, find $r^* \in \mathcal{R}^{\familymc}$ with \vspace{-0.5em} $$r^* \colonequals \arg\!\!\max\limits_{r\in \mathcal{R}_{\familymc}} \{\texttt{Prob}(D_{r}, \lozenge G) \mid D_r \models \Phi \mbox{ and } c(r) \leq B \}.$$
\end{enumerate}
The problem in feasibility synthesis is to determine a realisation satisfying all $\varphi \in \Phi$, or return that no such realisation exists. This problem is NP-complete~\cite{DBLP:conf/tacas/CeskaJJK19}.
The problem in max synthesis is to find a realisation that maximises the reachability probability of reaching $G$.
It can analogously be defined for minimising such probabilities.
As families are finite, such optimal realisations $r^*$ always exist.
It is beneficial to consider a variant of the max-synthesis problem in which the realisation $r^*$ is not required to achieve the maximal reachability probability, but it suffices to be  close to it.
This notion of \emph{$\varepsilon$-maximal synthesis} for a given $0< \varepsilon \leq 1$ amounts to find a realisation $r^*$ with $\texttt{Prob}(D_{r^*}, \lozenge G) \geq (1{-}\varepsilon) \cdot \max\limits_{r\in \mathcal{R}^{\familymc}} \{\texttt{Prob}(D_{r}, \phi)\}$.

\smallskip\noindent\emph{Problem statement and structure.}
In this paper, we propose novel synthesis algorithms for the probabilistic systems that are based on two concepts, CEGIS~\cite{DBLP:conf/pldi/Solar-LezamaRBE05} and syntax-guided synthesis~\cite{sygus}. 
To simplify the presentation, we start with CEGIS in Sect.~\ref{sec:CEGIS} and adopt it for MCs and the feasibility problem. 
In Sect.~\ref{sec:highlevel}, we lift and tune CEGIS, in particular towards probabilistic program sketches.

%% file: pics/realisation_2_2.tex
\begin{tikzpicture}[every node/.style={circle}]
	\node[draw] (1) {$0$};
	\node[draw, right=1.2 cm of 1] (0) {$1$} ;
	\node[draw, right=1.2 cm of 0] (2) {$2$};
	\node[draw=gray, right=1.2 cm of 2] (3) {\color{gray}$3$};
	\node[draw, right=1.2 cm of 3] (4) {$4$};

	\draw[->] (0) edge[bend right=60] node[above] {$0.1$} (1);
	
	\draw[->] (0) edge node[pos=0.75,above] {$0.8$} (2);
	\draw[->] (0) edge[bend left=30] node[pos=0.1,above] {$0.1$} (4);
	
	\draw[->] (1) edge[] node[above] {$0.5$} (0);
	\draw[->] (1) edge[bend right=20] node[pos=0.1,below] {$0.5$} (2);
	\draw[->] (2) edge[loop right] node[auto] {$1$} (2);

	\draw[->] (4) edge[loop above] node[auto] {$1$} (4);
	

	\draw[->,gray] (3) edge[loop below] node[left] {$1$} (3);

	\draw ($(1.north) + (0,0.3)$) edge[->] (1);
	\draw  [use as bounding box, draw=white] (-0.2,-1) rectangle (8.1,1.3) {};
\end{tikzpicture}%

%% file: pics/realisation_2_4.tex
\begin{tikzpicture}[every node/.style={circle}]
	\node[draw] (1) {$0$};
	\node[draw, right=1.2 cm of 1] (0) {$1$} ;
	\node[draw, right=1.2 cm of 0] (2) {$2$};

	\node[draw=gray, right=1.2 cm of 2] (3) {\color{gray}$3$};
	\node[draw, right=1.2 cm of 3] (4) {$4$};

	\draw[->] (0) edge[bend right=60] node[above] {$0.1$} (1);
	\draw[->] (0) edge[bend left=30] node[pos=0.1,above] {$0.9$} (4);

	\draw[->] (1) edge[] node[above] {$0.5$} (0);
	\draw[->] (1) edge[bend right=20] node[pos=0.1,below] {$0.5$} (2);

	\draw[->] (4) edge[loop above] node[auto] {$1$} (4);
	
	\draw[->] (2) edge[bend left=20] node[below, near start] {$1$} (4);

	\draw[->,gray] (3) edge[loop below] node[left] {$1$} (3);

	\draw ($(1.north) + (0,0.3)$) edge[->] (1);
	\draw  [use as bounding box, draw=white] (-1,-1) rectangle (7.2,1.3) {};
\end{tikzpicture}%

%% file: pics/realisation_3_2.tex
\begin{tikzpicture}[every node/.style={circle}]
	\node[draw] (1) {$0$};
	\node[draw, right=1.2 cm of 1] (0) {$1$} ;
	\node[draw, right=1.2 cm of 0] (2) {$2$};

	\node[draw, right=1.2 cm of 2] (3) {$3$};
	\node[draw, right=1.2 cm of 3] (4) {$4$};

	
	\draw[->] (0) edge[bend right=60] node[above] {$0.1$} (1);
	
	\draw[->] (0) edge node[pos=0.75,above] {$0.8$} (2);
	\draw[->] (0) edge[bend left=30] node[pos=0.1,above] {$0.1$} (4);
	
	\draw[->] (1) edge[] node[above] {$0.5$} (0);
	\draw[->] (2) edge[loop right] node[auto] {$1$} (2);

	\draw[->] (4) edge[loop above] node[auto] {$1$} (4);
	\draw[->] (1) edge[bend right=20] node[pos=0.1,below] {$0.5$} (3);
	

	\draw[->] (3) edge[loop below] node[left] {$1$} (3);

	\draw ($(1.north) + (0,0.3)$) edge[->] (1);
	\draw  [use as bounding box, draw=white] (-0.2,-1) rectangle (8.1,1.3) {};
\end{tikzpicture}%

%% file: pics/realisation_3_4.tex
\begin{tikzpicture}[every node/.style={circle}]
	\node[draw] (1) {$0$};
	\node[draw, right=1.2 cm of 1] (0) {$1$} ;
	\node[draw=gray,right=1.2 cm of 0] (2) {\color{gray}$2$};
	\node[draw, right=1.2 cm of 2] (3) {$3$};
	\node[draw, right=1.2 cm of 3] (4) {$4$};

	
	\draw[->] (0) edge[bend right=60] node[above] {$0.1$} (1);
	
	\draw[->] (0) edge[bend left=30] node[pos=0.1,above] {$0.9$} (4);
	\draw[->] (1) edge[] node[above] {$0.5$} (0);

	\draw[->] (4) edge[loop above] node[auto] {$1$} (4);
	\draw[->] (1) edge[bend right=20] node[pos=0.1,below] {$0.5$} (3);
	
	\draw[->,gray] (2) edge[bend left=20] node[above, near start] {$1$} (4);

	\draw[->] (3) edge[loop below] node[left] {$1$} (3);

	\draw ($(1.north) + (0,0.3)$) edge[->] (1);
	\draw  [use as bounding box, draw=white] (-1,-1) rectangle (7.2,1.3) {};
\end{tikzpicture}%

%% file: 03-CEGIS.tex
We follow the typical separation of concerns as in oracle-guided inductive synthesis~\cite{DBLP:journals/cacm/AlurSFS18,DBLP:conf/kbse/GerasimouTC15,DBLP:journals/ftpl/GulwaniPS17}:
a \emph{synthesiser} selects single realisations  $r$ that have not been considered before, and a \emph{verifier} checks whether the MC $D_r$ satisfies the specification $\Phi$ (cf.\ Fig.~\ref{fig:CEGIS} on page~\ref{fig:CEGIS}). 
If a realisation violates the specification, the verifier returns a \emph{conflict} representing the core part of the MC causing the violation. 

\subsection{Conflicts and synthesiser}
\label{sec:CEGIS:synthesiser}

To formalise conflicts, 
a \emph{partial realisation} of a family $\familymc$ is a function $\bar{r} \colon K \rightarrow S \cup \{\bot\}$ such that $\forall k\in K.~\bar{r}(k) \in T_k \cup {\{ \perp \}}$. 
For any partial realisations $\bar{r}_1$, $\bar{r}_2$, let $\bar{r}_1 \subseteq \bar{r}_2$ iff $\bar{r}_1(k) \in \{ \bar{r}_2(k), \bot \}$ for all $k \in K$. 

\begin{definition}[Conflict]
Let $r \in \mathcal{R}^{\familymc}$ be a realisation with $D_r \not\models \varphi$ for $\varphi \in \Phi$. A  partial realisation $\bar{r}_{\varphi} \subseteq r$ is a \emph{conflict} for the property $\varphi$ iff $D_{r'} \not\models \varphi$ for each realisation $r' \supseteq \bar{r}_{\varphi}$. 
A set of conflicts is called a \emph{conflict set}.
\end{definition}
To explore all realisations, the synthesiser starts with $\queue \colonequals \mathcal{R}^\familymc$ and picks some realisation $r \in \queue$.\footnote{We focus on program-level synthesis, and refrain from discussing important implementation aspects---like how to represent $\queue$---here.} 
Either $D_r \models \Phi$ and we immediately return $r$, or a conflict is found:
then $\queue$ is pruned by removing all conflicts that the verifier found.
If $\queue$ is empty, we are done: each realisation violates a property $\varphi \in \Phi$.

\subsection{Verifier}
\label{sec:CEGIS:verifier}

\begin{definition}
\label{def:verifier}
  A verifier is sound and complete, if for family $\familymc$, realisation $r$, and specification $\Phi$, the verifier terminates, the returned conflict set is empty iff $D_r \models \Phi$, and if it is not empty, it contains a conflict $\bar{r}_{\varphi} \subseteq r$ for some $\varphi \in \Phi$.
\end{definition}
Algorithm~\ref{alg:ver} outlines a basic verifier. It uses an off-the-shelf probabilistic model-checking procedure \textsc{Check($D_r$, $\varphi$)} to determine  which $\varphi \in \Phi$ (if any) are violated.
The algorithm then iterates over the violated $\varphi$ and computes critical sets $C$ of  $D_r$ that induce sub-MCs such that $\submc{D_r}{C}\not\models \varphi$ (line 6). The critical  sets for  safety properties can be obtained via standard methods~\cite{Abraham2014}, support for liveness properties is discussed at the end of the section.

\begin{example}
\label{ex:cegisfamily}
Reconsider $\familymc$ from Ex.~\ref{ex:dtmc_family} with $\Phi \colonequals \{ \varphi \colonequals \mathbb{P}_{\leq \nicefrac{2}{5}}(\lozenge \{ 2 \}) \}$.
Assume the synthesiser picks realisation $r_1$.
The verifier builds $D_{r_1}$ and determines $D_{r_1} \not\models \Phi$.
Observe that the verifier does not need the full realisation $D_{r_1}$ to refute~$\Phi$.
In fact, the paths in the fragment of $D_{r_1}$ in Fig.~\ref{fig:fragment} 
(ignoring the outgoing transitions of states $1$ and $2$) 
suffice to show that the probability to reach state $2$ exceeds $\nicefrac 2 5$.
Formally, the fragment in Fig.~\ref{fig:submodel} is a sub-MC $\submc{D_{r_1}}{C}$ with critical states $C = \{ 0 \}$.
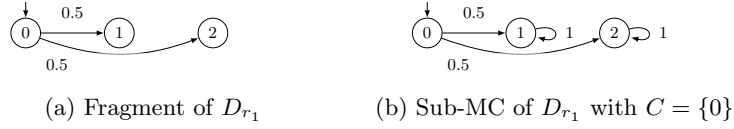
\begin{figure}[t]
\centering
\subfloat[Fragment of $D_{r_1}$]{
\scalebox{0.7}{\input{pics/submc}}
	\label{fig:fragment}
}
	\subfloat[Sub-MC of $D_{r_1}$ with $C=\{0\}$]{
	\scalebox{0.7}{\input{pics/submcproper}}
	\label{fig:submodel}
	}
	\caption{Fragment and corresponding sub-MC that suffices to refute $\Phi$}
\end{figure} 
The essential property is~\cite{DBLP:conf/tacas/WimmerJABK12}:
\[
\mbox{\emph{If a sub-MC of a MC $D$ refutes a safety property $\varphi$, then $D$ refutes $\varphi$ too.}}
\]
Observe that $\submc{D_{r_1}}{C}$ is part of $D_{r_2}$ too. 
Formally, the sub-MC of $\submc{D_{r_2}}{C}$ is isomorphic to $\submc{D_{r_1}}{C}$ and therefore also violates $\Phi$.
Thus, $D_{r_2} \not\models \Phi$.
\end{example}
Finally, the verifier translates the obtained critical set $C$ for realisation $r$ to a conflict $\textsl{Conflict}(C,r) \subseteq r$ and stores it in the conflict set $\mathsf{Conflict}$ (line~7). 
The procedure \textsc{generateConflict}$(\familymc,r,C)$ identifies the subset of parameters~$K$ that occur in the sub-MCs $\submc{D_r}{C}$ and returns the corresponding partial realisation. 
The proposition below clarifies the relation between critical sets and conflicts.

\begin{algorithm}[t]
\caption{Verifier}
\label{alg:ver}
\begin{footnotesize}
\renewcommand{\baselinestretch}{1}
\begin{algorithmic}[1]
\Function{Verify}{family $\familymc$, realisation $r$, specification $\Phi$}
\State $\mathsf{Violated} \gets \emptyset$; $\mathsf{Conflict} \gets \emptyset$; $D_r \gets \textsc{GenerateMC}(\familymc, r)$;
  \ForAll{$\varphi \in \Phi$}
  \IIf {not \textsc{Check}$(D_r, \varphi)$} $\mathsf{Violated} \gets \mathsf{Violated} \cup \{ \varphi \}$ \EndIIf 
  \EndFor
  \ForAll{$\varphi \in  \mathsf{Violated}$}
  	\State $C_\varphi \gets$ \textsc{ComputeCriticalSet}$(D_r, \varphi)$ 
  \State $\mathsf{Conflict} \gets \mathsf{Conflict}~\cup $ \textsc{generateConflict}$(\familymc,r,C_\varphi)$
  \EndFor
  \State \Return $\mathsf{Conflict}$
\EndFunction
\end{algorithmic}
\end{footnotesize}
\end{algorithm}

\begin{proposition}
\label{prop:counterexgeneralisation}
	If $C$ is a critical set for $D_r$ and $\varphi$, then $C$ is also a critical set for each $D_{r'}$, $r' \supseteq \textsl{Conflict}(C,r)$, and $D_{r'} \not\models \varphi$.
\end{proposition}

\begin{example}
Recall from Ex.~\ref{ex:cegisfamily} that $D_{r_2} \not\models \Phi$.
This can be concluded \emph{without constructing $D_{r_2}$}.
Just considering $r_2$, $\familymc$ and $C$ suffices:
First, take all parameters occurring in $\probfam(c)$ for any $c \in C$.
This yields $\{ k_1, k_2 \}$. The partial realisation $\bar{r} \colonequals \{ k_1 \mapsto 1, k_2 \mapsto 2 \}$ is a conflict.
The values for the other parameters do not affect the shape of the sub-MC induced by $C$.
Realisation $r_2 \supset \bar{r}$ only varies from $r_1$ in the value of $k_3$, but $\bar{r}(k_3) = \bot$, i.e., $k_3$ is not included in the conflict.
This suffices to conclude $D_{r_2} \not\models \Phi$.
\end{example}

\smallskip\noindent\textbf{Conflicts for liveness properties.}
To support liveness properties such as $\varphi \colonequals \mathbb{P}_{> \lambda}(\lozenge G)$, we first consider a (standard) dual safety property $\varphi' \colonequals \mathbb{P}_{< 1-\lambda}(\lozenge B)$, where $B$ is the set of all states that do not have a path to $G$.
Observe that $B$ can be efficiently computed using graph algorithms.
We have to be careful, however. 
\begin{example}
	Consider $D_{r_1}$, and let $\varphi \colonequals \mathbb{P}_{> 0.6}(\lozenge \{4\})$. $D_{r_1} \not\models \varphi$. 
	Then, $\varphi' = \mathbb{P}_{< 0.4}(\lozenge \{2\})$, which is refuted with critical set $C=\{ 0 \}$ as before.
	Although $\submc{D_{r_2}}{C}$ is again isomorphic to $\submc{D_{r_1}}{C}$, we have $D_{r_2} \models \varphi$. 
	The problem here is that state $2$ is in $B$ for $D_{r_1}$ as $r_1(k_3) = 2$, but not in $B$ for $D_{r_2}$, as $r_2(k_3) = 4$.   
	
\end{example}
To prevent the problem above, we ensure that the states in $B$ cannot reach $G$ in other realisations, by including $B$ in the critical set of $\varphi$: 
Let $C$ be the critical set for the dual safety property $\varphi'$. We define $B \cup C$ as critical states for $\varphi$. 
Together, we reach states $B$ with a critical probability mass\footnote{A good implementation takes a subset of $B' \subseteq B$ by considering the  $\texttt{Prob}(D,\lozenge B')$.}, and never leave $B$.
\begin{example}
	In $D_{r_1}$, we compute critical states $\{ 0, 2 \}$, preventing the erroneous reasoning from the previous example.
	For $D_{r_4}$, we compute $C'=\{ 0 \} \cup \{ 3 \}$ as critical states, and as $\submc{D_{r_4}}{C'}$ is isomorphic to $\submc{D_{r_3}}{C'}$, we obtain that $D_{r_3} \not\models \varphi$.
\end{example}

%% file: pics/submc.tex
\begin{tikzpicture}[every node/.style={circle}]
	\node[draw] (1) {$0$};
	\node[draw, right=1.2 cm of 1] (0) {$1$} ;
	\node[draw, right=1.2 cm of 0] (2) {$2$};

	\draw[->] (1) edge[] node[above] {$0.5$} (0);
	\draw[->] (1) edge[bend right=20] node[pos=0.1,below] {$0.5$} (2);

	


	\draw ($(1.north) + (0,0.3)$) edge[->] (1);
	\draw  [use as bounding box, draw=white] (-1.2,-0.4) rectangle (6.1,0.8) {};
\end{tikzpicture}%

%% file: pics/submcproper.tex
\begin{tikzpicture}[every node/.style={circle}]
	\node[draw] (1) {$0$};
	\node[draw, right=1.2 cm of 1] (0) {$1$} ;
	\node[draw, right=1.2 cm of 0] (2) {$2$};

	\draw[->] (0) edge[loop right] node[auto] {$1$} (0);
	\draw[->] (2) edge[loop right] node[auto] {$1$} (2);
	
	\draw[->] (1) edge[] node[above] {$0.5$} (0);
	\draw[->] (1) edge[bend right=20] node[pos=0.1,below] {$0.5$} (2);
	

	


	\draw ($(1.north) + (0,0.3)$) edge[->] (1);
	\draw  [use as bounding box, draw=white] (-1.2,-0.4) rectangle (6.1,0.8) {};
\end{tikzpicture}%

%% file: 03-approach.tex
\section{Syntax-Guided Synthesis for Probabilistic Programs}
\label{sec:highlevel}

Probabilistic models are typically specified by means of a program-level modelling language, such as PRISM~\cite{KNP11}, PIOA~\cite{DBLP:journals/tcs/WuSS97}, JANI~\cite{DBLP:conf/tacas/BuddeDHHJT17}, or MODEST~\cite{DBLP:journals/tse/BohnenkampDHK06}. We propose a \emph{sketching language} based on the PRISM modelling language. A sketch, a syntactic template, defines a high-level structure of the model and represents a-priori knowledge about the system under development. It effectively restricts the size of the design space and also allows to concisely add constraints and costs to its members. The proposed language is easily supported by model checkers and in particular by methods for generating CEs~\cite{DBLP:conf/atva/DehnertJWAK14,DBLP:journals/corr/abs-1305-5055}. 
Below, we describe the language, and adapt CEGIS from state level to program level. In particular, we employ so-called \emph{program-level CEs}, rather than CEs on the state level.

\subsection{A program sketching language}
Let us briefly recap how the model-based concepts translate to language concepts in  
the PRISM guarded-command language.
A PRISM program consists of one or more reactive modules that may interact with each other. 
Consider a single module.
This is not a restriction, every PRISM program can be flattened into this form. 
A module has a set of bounded variables spanning its state space.
Transitions between states are described by guarded commands of the form:
$$
\mbox{\texttt{guard}} 
\ \ \rightarrow \ \ 
p_1 : \mbox{\texttt{update}}_1 + \ldots \ldots + p_n : \mbox{\texttt{update}}_n 
$$
The guard is a Boolean expression over the module's  variables of the model. 
If the guard evaluates to true, the module can evolve into a successor state by updating its variables. 
An update is chosen according to the probability distribution given by expressions $p_1, \hdots, p_n$.
In every state enabling the guard, the evaluation of $p_1,\hdots,p_n$ must sum up to one.
Overlapping guards yield non-determinism and are disallowed here.
Roughly, a program $\program{P}{}$ thus is a tuple $(\Vars, E)$ of variables and commands. 
	For a program $\program{P}{}$, the \emph{underlying MC} $\semantics{\program{P}{}}$ are $\program{P}{}$'s semantics.
We lift specifications: Program $\program{P}{}$ satisfies a specification $\Phi$, iff $\semantics{\program{P}{}} \models \Phi$, etc.

A sketch is a program that contains \emph{holes}. 
Holes are the program's open parts and can be replaced by one of finitely many options. Each option can \emph{optionally} be named and associated with a cost.
They are declared as:
$$
\mbox{\texttt{hole} } h \mbox{ \texttt{either} }  
\{ \, x_1 \mbox{\texttt{ is }} \mbox{\texttt{expr}}_1 \mbox{\texttt{ cost }} c_1, 
\ldots, 
x_k \mbox{\texttt{ is }} \mbox{\texttt{expr}}_k \mbox{\texttt{ cost }} c_k
\, \}
$$
where $h$ is the hole identifier, $x_i$ is the option name, $\mbox{\texttt{expr}}_i$ is an expression over the program variables describing the option, and $c_i$ are the cost, given as expressions over natural numbers.
A hole $h$ can be used in commands in a similar way as a constant, and may occur multiple times within multiple commands, in both guards and updates. 
The option names can be used to describe constraints on realisations.
These propositional formulae over option names restrict realisations,~e.g.,
$$
\mbox{\texttt{constraint} } (x_1 \, \vee \, x_2) \, \Longrightarrow x_3
$$
requires that whenever the options $x_1$ or $x_2$ are taken for some (potentially different) holes, option $x_3$ is also to be taken.

\begin{definition}[Program sketch]
A \emph{(PRISM program) sketch} is a pair 
$\Sketch \colonequals (\program{P}{H}, \Option_H, \SketchConstraints, \OptionPrices)$
 where $\program{P}{H}$ is a program with a set $H$ of holes with options $\Option_H$, $\SketchConstraints$ are constraints over $\Option_H$, and $\OptionPrices\colon \Option_H \rightarrow \mathbb{N}$ option-costs.
\end{definition}	
\lstset{language=Prism}   
\newsavebox{\sketch}
\begin{lrbox}{\sketch}
\begin{lstlisting}
hole X either { XA is 1 cost 3, 2}
hole Y either { YA is 1, 3 }
hole Z either { 1, 2 }
constraint  !(XA && YA);
module rex
s : [0..3] init 0;
s = 0 -> 0.5: s'=X + 0.5: s'=Y;
s = 1 -> s'=s+Z; 
s >= 2 -> s'=s;
endmodule
\end{lstlisting}
\end{lrbox}
\newsavebox{\instance}
\begin{lrbox}{\instance}
\begin{lstlisting}
module rex
s : [0..3] init 0;
s = 0 -> 0.5: s'=1 + 0.5: s'=3;      
s = 1 -> s'=3; 
s >= 2 -> s'=s;
endmodule
\end{lstlisting}
\end{lrbox}

\begin{figure}[t]
\subfloat[Program sketch $\Sketch$]{\usebox{\sketch}
\label{fig:runningexample:sketch}
} \hspace{2em}%
\subfloat[Instance $\Sketch {(\{ X{\mapsto}1, Z{\mapsto}2, Y{\mapsto}3 \})}$]{\usebox{\instance}\hspace{3em}
\label{fig:runningexample:instance}
}
\caption{Running example}
\end{figure}
\begin{example}
\label{ex:runningex:sketch}
\label{ex:runningex:dtmcs}
We consider a small running example to illustrate the main concepts.
Fig.~\ref{fig:runningexample:sketch} depicts the program sketch $\Sketch$ with 
holes $H = \{ \text{X}, \text{Y}, \text{Z} \}$.
For \text{X}, the options are $\Option_\text{X} = \{ 1, 2 \}$.
The constraint forbids \text{XA} and \text{XB} both being one; it ensures a non-trivial random choice in state \texttt{s=0}.
\end{example}

\begin{remark}
	Below, we formalise notions previously used on 
	families. 
	Due to flexibility of sketching (in particular in combination with multiple modules), it is \emph{not} straightforward to provide family semantics to sketches, but the concepts are analogous. In particular: holes and parameters are similar, parameter domains are options, and family realisations and sketch realisations both yield concrete instances from a family/sketch. The synthesis problems carry over naturally.
\end{remark}
\begin{definition}[Realisations of sketches]
\label{def:realisation}
Let $\Sketch \colonequals (\program{P}{H}, \Option_H, \SketchConstraints, \OptionPrices)$ be a sketch, a \emph{sketch realisation} on holes $H$ is a function $R \colon H \rightarrow \Option_H$ with $\forall h \in H.~R(h) \in \Option_h$ and that satisfies all constraints in $\SketchConstraints$.
The \emph{sketch instance} $\Sketch(R)$ for realisation $R$ is the program (without holes) $\program{P}{H}[H / R]$ in which each hole $h \in H$ in $\program{P}{H}$ is replaced by $R(h)$.
The cost $c(R)$ is the sum of the 
cost of the selected options, $c(R) \colonequals \sum_{h \in H} \OptionPrices(R(h))$.
\end{definition}

\begin{example}
\label{ex:extrarealisations}
We continue Ex.~\ref{ex:runningex:sketch}.
The program in Fig.~\ref{fig:runningexample:instance} reflects $\Sketch(R)$ for realisation $R = \{ X {\mapsto} 1, Z {\mapsto} 2, Y {\mapsto} 3\}$, with $c(R) = 3$ as $\OptionPrices(R(X)) = 3$  and all other options have cost zero.
For realisation $R' = \{ Y, Z \mapsto 1, X \mapsto 2 \}$, 
$c(R') = 0$.
The assignment $\{ X, Y, Z \mapsto 1\}$ violates the constraint and is not a realisation.
In total, $\Sketch$ represents $6 = 2^3{-}2$ programs and their underlying MCs.
\end{example}

%% file: synthesiser.tex
\subsubsection*{Feasibility synthesis.}
The synthesiser follows the steps in Alg.~\ref{alg:syn}. 
During the synthesis process, the synthesiser stores and queries the set of realisations not yet pruned. 
These remaining realisations are represented by (the satisfying assignments of) the first-order formula $\psi$ over hole-assignments.
Iteratively extending $\psi$ with conjunctions thus prunes the remaining design space.

\begin{algorithm}[t]
\caption{Synthesiser (feasibility synthesis)}
\label{alg:syn}
\begin{footnotesize}
\renewcommand{\baselinestretch}{1}
\begin{algorithmic}[1]
\Function{Synthesis}{program sketch $\Sketch$, specification $\Phi$, budget $B$}
  \State $\psi \gets$ \textsc{Initialise}$(\Sketch, B)$ 
    \State $R \gets$ \textsc{GetRealisation}$(\psi)$
  \While{$R \neq$  \texttt{Unsat}} 
     \State $C \gets$ \textsc{Verify}$(\Sketch(R), \Phi)$
          \IIf{$C = \emptyset$} \textbf{return}  $R$ \EndIIf

      \State $\psi \gets \psi \wedge \Big(\bigwedge_{\bar{R} \in C}$ \textsc{LearnFromConflict}$(\Sketch,\bar{R})\Big)$
       \State $R \gets$ \textsc{GetRealisation}$(\psi)$ 
         \EndWhile
\State \textbf{return}  \texttt{Unsat}
\EndFunction
\end{algorithmic}
\end{footnotesize}
\renewcommand{\baselinestretch}{1}
\end{algorithm}
We give a brief overview, before detailing the steps.
\textsc{Initialise}$(\Sketch, B)$ constructs $\psi$ such that it represents \emph{all} sketch realisations that satisfy the constraints in the sketch $\Sketch$ within the budget $B$.
\textsc{GetRealisation($\psi$)} exploits an SMT-solver for linear (bounded) integer arithmetic to obtain a realisation $R$ consistent with $\psi$, or \texttt{Unsat} if no such realisation exists.
As long as new realisations are found, the verifier analyses them (line 5) and returns a conflict set $C$.
If $C = \emptyset$, the $\Sketch(R)$ satisfies the specification $\Phi$ and the search is terminated.
Otherwise, the synthesiser updates $\psi$ based on the  conflicts (line 7).
$R$ is always pruned.

\smallskip\noindent{\textsc{Initialise}$(\Sketch, B)$}: 
Let hole $h \in H$ have (ordered) options $\Option_h = \{ o_h^1, \hdots, o_h^n \}$. 
To encode realisation $R$, we introduce integer-valued meta-variables $K_H \colonequals \{ \kappa_h \mid h \in H\}$ with the semantics that $\kappa_h = i$ whenever hole $h$ has value $o_h^i$, i.e., $R(h) = o_h^i$.
We set $\psi \colonequals \psi_\text{opti} \land \psi_\Gamma \land \psi_\text{cost}$, where $\psi_\text{opti}$ ensures that each hole is assigned to some option, $\psi_\Gamma$ ensures that the sketch's constraints $\Gamma$ are satisfied, and $\psi_\text{cost}$ ensures that the budget is respected.
These sub-formulae are:
\begin{align*}
&\psi_\text{opti} \colonequals \bigwedge_{h \in H} 1 \leq \kappa_h \leq |\Option_h|,\quad \ \ \psi_\Gamma \colonequals \bigwedge_{\gamma \in \Gamma} \gamma[N_h^i/\kappa_h = i],\\
&\psi_\text{cost} \colonequals \sum_{h \in H} \omega_h \leq B \land \left( \bigwedge_{h \in H} \bigwedge_{i=1}^n \kappa_h = i \rightarrow \omega_h = \OptionPrices(o_h^i) \right),
\end{align*}
where  $\gamma[N_h^i/\kappa_h = i]$ denotes that in every constraint $\gamma \in \Gamma$ we replace each option name $N_h^i$ for an option $o_h^i$ with $\kappa_h = i$, and $\omega_h$ are fresh variables storing the cost for the selected option at hole $h$.

\begin{example}
\label{ex:runningexample:psi0}
For sketch $\Sketch$ in Fig.~\ref{fig:runningexample:sketch}, we obtain (with slight simplifications) \begin{align*}
 & \psi \colonequals
 1 \leq \kappa_X \leq 2 \land 1 \leq \kappa_Y \leq 2 \land 1 \leq \kappa_Z \leq 2 \land \neg (\kappa_X = 1 \land \kappa_Y = 1) \land \\
   & \omega_X + \omega_Y + \omega_Z \leq B \land \kappa_X = 1 \rightarrow \omega_X = 3 \land \kappa_X = 2 \rightarrow \omega_X = 0 \land \omega_Y = 0 \land \omega_Z = 0.	
 \end{align*}
\end{example}

\smallskip\noindent{\textsc{GetRealisation}$(\psi)$}:
To obtain a realisation $R$, we check satisfiability of $\psi$. 
The solver either returns \texttt{Unsat} indicating that the synthesiser is finished, or \texttt{Sat}, together with a satisfying assignment $\alpha_{R} \colon K_H \rightarrow \mathbb{N}$.
The assignment $\alpha_R$ uniquely identifies a realisation $R$ by $R(h) \colonequals o_h^{\alpha_R(\kappa_h)}$. The sum over $\omega_{*}$ gives $c(R)$.


\smallskip\noindent{\textsc{Verify}$(\Sketch(r), \Phi)$}: invokes any sound and complete verifier, e.g., an adaption of the verifier from Sect.~\ref{sec:CEGIS:verifier} as presented in Sect.~\ref{sec:highlevel:verifier}.

\smallskip\noindent{\textsc{LearnFromConflict}$(\Sketch,\bar{R})$}:
For a conflict
\footnote{As in Sect.~\ref{sec:CEGIS:synthesiser}: 
A \emph{partial realisation} for $\Sketch$ is a function ${\bar{R} \colon H \rightarrow \Option_H \cup \{\bot\}}$ s.t.\ $\forall h\in H.~\bar{R}(h) \in \Option_h \cup {\{ \perp \}}$. 
For partial realisations $\bar{R}_1, \bar{R}_2$, let $\bar{R}_1 \subseteq \bar{R}_2$ iff $\forall h \in H.\;\bar{R}_1(h) \in \{ \bar{R}_2(h),\perp\}$. 
Let $R$ be a realisation s.t.\ $\Sketch(R) \not\models \varphi$ for $\varphi \in \Phi$. 
Partial realisation $\bar{R}_{\varphi} \subseteq R$ is a \emph{conflict} for $\varphi$ iff $\forall R' \supseteq \bar{R}_{\varphi}~\Sketch(R')  \not\models \varphi$. 
} $\bar{R} \in C$, we add the formula  \[ \neg \Big( \bigwedge_{h \in H, \bar{R}(h) \neq \perp} \kappa_h = \alpha_{\bar{R}}(\kappa_h) \Big). \]
that excludes realisations $R' \supseteq \bar{R}$. Intuitively, the formula states that the realisations remaining in the design space (encoded by the updated $\psi$) must have different valuations of holes $h$ w.r.t.\ $\bar{R}$ (for holes where $\bar{R}(h) \neq \bot$).

\begin{example}
\label{ex:runningex:alpha0}
Consider $\psi$ from Ex.~\ref{ex:runningexample:psi0}.
The satisfying assignment (for $B \geq 2$) is $\{ \kappa_X \mapsto 1, \kappa_Y, \kappa_Z \mapsto 2, \omega_X \mapsto 3, \omega_Y, \omega_Z \mapsto 0 \}$ represents $R$, $c(R) = 3$ from Ex.~\ref{ex:runningex:sketch}.
Consider $\Phi = \{\mathbb{P}_{\leq 0.4} [\lozenge~\texttt{s=3}] \}$.
The verifier (for now, magically) constructs a conflict set $\{ \bar{R} \}$ with $\bar{R} = \{ \text{Y} \mapsto 3 \}$.
The synthesiser updates $\psi \gets \psi \land \kappa_Y \neq 2$ (recall that $\kappa_Y = 2$ encodes $Y \mapsto 3$).
A satisfying assignment $\{ \kappa_X, \kappa_Y, \kappa_Z \mapsto 1 \}$ for $\psi$ encodes $R'$ from Ex.~\ref{ex:extrarealisations}.
As $\Sketch(R') \models \Phi$, the verifier reports no conflict.
\end{example}

\medskip\noindent\textbf{Optimal synthesis.}
We adapt the synthesiser to support max synthesis, cf.\ Alg.~\ref{alg:opt}.
\begin{algorithm}[t]
\caption{Synthesiser (max synthesis)}
\label{alg:opt}
\begin{footnotesize}
\renewcommand{\baselinestretch}{1}
\begin{algorithmic}[1]
\Function{Synthesis}{$\Sketch$, $\Phi$, $B$, goal predicate $G$, tolerance $\varepsilon$}
  \State $\lambda^{*} \gets \infty$, $R^* \gets\texttt{Unsat}$,  $\psi \gets$ \textsc{Initialise}$(\Sketch, B)$ 
    \State $R \gets$ \textsc{getRealisation}$(\psi)$
  \While{$R \neq$  \texttt{Unsat}} 
     \State $C, \lambda_\text{new} \gets$ \textsc{OptimiseVerify}$(\Sketch(R), \Phi, G, \lambda^{*}, \varepsilon)$ \label{alg:optimising_synthesiser_optimiseverify}
          \IIf{ $C = \emptyset$} $\lambda^{*}, R^* \gets \lambda_\text{new}, R$ \EndIIf
      \State $\psi \gets \psi \wedge \Big(\bigwedge_{\bar{R} \in C}$ \textsc{LearnFromConflict}$(\Sketch,\bar{R})\Big)$
       \State $R \gets$ \textsc{getRealisation}$(\psi)$ 
         \EndWhile
\State \textbf{return} $R^*$
\EndFunction
\end{algorithmic}
\end{footnotesize}
\end{algorithm} 
Recall the problem aims at maximizing the probability of reaching states described by a predicate $G$, w.r.t.\ the tolerance $\varepsilon \in (0,1)$.
Algorithm~\ref{alg:opt} stores in $\lambda^{*}$ the maximal probability $\Prob(\Sketch(R), \lozenge G)$ 
among all considered realisations $R$, and this $R$ in $R^*$. 
In each iteration, an optimising verifier is invoked (line~\ref{alg:optimising_synthesiser_optimiseverify}) on realisation $R$. 
If $\Sketch(R) \models \Phi$ and $\Prob(\Sketch(R), \lozenge G)\;{>}\; \lambda^{*}$, it returns an empty conflict set \emph{and} $\lambda_\text{new} \colonequals \Prob(\Sketch(R), \lozenge G)$.
Otherwise, it reports a conflict set for $\Phi \cup \{ \mathbb{P}_{\geq (1-\varepsilon) \cdot \lambda^{*}}(\lozenge G)\}$.%

%% file: highlevelverifier.tex
We now adapt the statel-level verifier from Sect.~\ref{sec:CEGIS:verifier} in Alg.~\ref{alg:ver} 
to use program-level counterexamples~\cite{DBLP:journals/corr/abs-1305-5055} for generating conflicts. 
The appendix contains more details. 

\smallskip\noindent\emph{\textsc{generateMC}$(\Sketch,R)$}:
This procedure first constructs the instance $\Sketch(R)$, i.e., a program without holes, from $\Sketch$ and $R$, as in Fig.~\ref{fig:runningexample:instance}: 
Constraints in the sketch are removed, as they are handled by the synthesiser.
This approach allows us to use any model checker supporting PRISM programs. 
The realisation is passed separately, the sketch is parsed \emph{once} and then  appropriately instantiated.
The instance is then translated into the underlying MC $\semantics{\Sketch(R)}$ via standard procedures, with transitions annotated with their generating commands.

\newsavebox{\instanceReal}
\begin{lrbox}{\instanceReal}
\begin{lstlisting}
const int X = 1;
const int Y = 3;
const int Z = 3;
module rex
s : [0..3] init 0;
s=0 -> 0.5:s'=X + 0.5:s'=Y;
s=1 -> s'=Z; 
s=2 -> s'=2;
s=3 -> s'=3;
endmodule
\end{lstlisting}
\end{lrbox}
\newsavebox{\highlevelCounterex}
\begin{lrbox}{\highlevelCounterex}
\begin{lstlisting}
const int X = 1, Y = 3;
...
module rex
s : [0..3] init 0;
s=0 -> 0.5: s'=X + 0.5: s'=Y;
endmodule
\end{lstlisting}
\end{lrbox}
\newsavebox{\branchsplitting}
\begin{lrbox}{\branchsplitting}
\begin{lstlisting}
...
module rex
s : [0..3] init 0;
t : [0..2] init 0;
s=0 & t=0 -> 0.5:t'=1 + 0.5:t'=2;
s=0 & t=1 -> 1:s'=X & t'=0;
s=0 & t=2 -> 1:s'=Y & t'=0;
...
endmodule
\end{lstlisting}
\end{lrbox}
\newsavebox{\lbcounterexample}
\begin{lrbox}{\lbcounterexample}
\begin{lstlisting}
...
module rex
s : [0..3] init 0;
s=0 -> 0.5:s'=X + 0.5:s'=Y;
s=3 -> s'=3
endmodule
\end{lstlisting}
\end{lrbox}

\begin{figure}[t]
\subfloat[CE for upper bound]{
\usebox{\highlevelCounterex}
\label{fig:runningex:counterex}
}
\hspace{3em}
\subfloat[CE for lower bound]{

\usebox{\lbcounterexample}

\label{fig:lowerboundcounterex}
}
\caption{CEs for (a) $\mathbb{P}_{\leq 0.4} [F~\texttt{s=3}]$ and (b) $\mathbb{P}_{>0.6} [F~\texttt{s=2}]$.}
\end{figure}

\smallskip\noindent{\textsc{ComputeCriticalSet}$(D, \varphi)$}  computes program-level CEs as analogue of critical sets. They are defined on commands rather than on states.
Let $\program{P}{} = (\Vars, E)$ be a program with commands $E$. 
Let ${\program{P}{}}_{|{E'}} \colonequals (\Vars, E')$ denote the restriction of $\program{P}{}$ to $E'$  (with variables and initial sates as in $\program{P}{}$).
Building ${\program{P}{}}_{|{E'}}$ 
 may introduce deadlocks in $\semantics{{\program{P}{}}_{|{E'}}}$ (just like a critical set introduces deadlocks). 
 To remedy this, we use the standard operation $\fixdeadlock$, which takes a program and adds commands that introduce self-loops for states without enabled guard.

\begin{definition}
\label{def:genCE}
	For program $\program{P}{} = (\Vars, E)$ and specification $\Phi$ with $\program{P}{} \not\models \Phi$, a program-level CE  $E' \subseteq E$ is a set of commands, such that for all (non-overlapping) programs $\program{P'}{} = (\Vars, E'')$ with $E'' \supseteq E'$ (i.e, extending $\program{P}{}$), $\fixdeadlock(\program{P'}{}) \not\models \Phi$.
\end{definition}

\begin{example}
\label{ex:runningex:overlapping}
Reconsider $\Phi =  \{ \mathbb{P}_{\leq 0.4} [\lozenge~\texttt{s=3}] \}$.
Figure~\ref{fig:runningex:counterex} shows a CE for $\Sketch(R)$ in Fig.~\ref{fig:runningexample:instance}.
The probability to reach \texttt{s=3} in the underlying MC is $0.5 > 0.4$. 
\end{example}
For safety properties, program-level CEs coincide with  high-level CEs proposed in~\cite{DBLP:journals/corr/abs-1305-5055}, their extension to liveness properties follows the ideas on families.
The program-level CEs are computed by an extension of the MaxSat~\cite{DBLP:series/faia/2009-185} approach from~\cite{DBLP:conf/atva/DehnertJWAK14}. The appendix contains details and further extensions.


\smallskip\noindent{\textsc{GenerateConflict}$(\Sketch,R,E)$} generates conflicts from commands: we map commands in $\Sketch(R)_{|E}$ to the commands from $\Sketch$, i.e., we restore the information about the critical holes corresponding to the part of the design space that can be pruned by CE $E$. Formally, $\textsl{Conflict}(E,R)(h) = R(h)$ for all $h \in H$ that appear in restriction $\Sketch{}_{|E}$.
\begin{proposition}
\label{prop:counterexgeneralisation}
	If $E$ is a CE for $\Sketch(R)$, then $E$ is also a CE for each $\Sketch(R')$,  $R' \supseteq \textsl{Conflict}(E,R)$.
\end{proposition}

\begin{example}
\label{ex:runningex:counterex}
The CEs in Fig.~\ref{fig:runningex:counterex} contain commands which depend on the realisations for holes X and Y.
For these fixed values, the program violates the specification \emph{independent of the value for Z}, so Z is not in the conflict
$\{ \text{X} \mapsto 1, \text{Y} \mapsto 3 \}$.
\end{example}

%% file: experiments.tex
\section{Experimental Evaluation and Discussion}

\noindent\emph{Implementation.}
We evaluate the synthesis framework with a prototype\footnote{\url{https://github.com/moves-rwth/sketching}} using the SMT-solver~Z3~\cite{Z3}, and (an extension of) the model checker Storm~\cite{DBLP:conf/cav/DehnertJK017}.

\medskip\noindent\textbf{Case studies.}
We consider the following three case studies:

\smallskip\noindent\textsl{Dynamic power management (DPM).}
The goal of this adapted DPM problem~\cite{BBPM99} is to trade-off power consumption 
for performance. 
We sketch a controller that decides based on the current workload
, inspired by~\cite{DBLP:conf/kbse/GerasimouTC15}. The fixed environment contains no holes.
The goal is to synthesise the guards and updates to satisfy a specification with properties such as $\varphi_1$: the expected number of lost requests is below $\lambda$, and $\varphi_2$: the expected energy consumption is below $\kappa$.

\smallskip\noindent\textsl{Intrusion} describes a network (adapted from~\cite{DBLP:journals/tcs/KwiatkowskaNPV09}), in which the controller tries to infect a target node via intermediate nodes.
A failed attack makes a node temporarily harder to intrude.
We sketched a partial strategy aiming to minimise the expected time to intrusion. 
Constraints encode domain specific knowledge.

\smallskip\noindent\textsl{Grid}
 is based on a classical benchmark for solving partially observable MDPs (POMDPs)~\cite{DBLP:journals/ai/KaelblingLC98}. To solve POMDPs, the task is to find an observation-based strategy, which is undecidable for the properties we consider. Therefore, we resort to finding a deterministic $k$-state strategy~\cite{meuleau1999solving} s.t.\ in expectation, the strategy requires less than $\lambda$ steps to the~target. This task is still hard:
finding a memoryless, observation-based strategy is already NP-hard \cite{VlassisLB12,DBLP:conf/aaai/ChatterjeeCD16}.
We create a family describing all $k$-state strategies (for some fixed $k$) for the POMDP. Like in \cite{DBLP:conf/uai/Junges0WQWK018} actions are reflected by parameters, while parameter dependencies ensure that the strategy is observation-based.

\medskip\noindent\textbf{Evaluation.} 
%
We compare w.r.t.\ an enumerative approach.
That baseline linearly depends on the number of realisations, and the underlying MCs' size. 
We focus on sketches where all realisations are explored, as relevant for optimal synthesis. For concise presentation we use $\texttt{Unsat}$ variants of feasibility synthesis, where methods perform mostly independent of the order of exploring realisations.
We evaluate results for \textsl{DPM}, and summarise further~results. All results are obtained on a Macbook~MF839LL/A, within 3 hours and using less than 8~GB RAM.

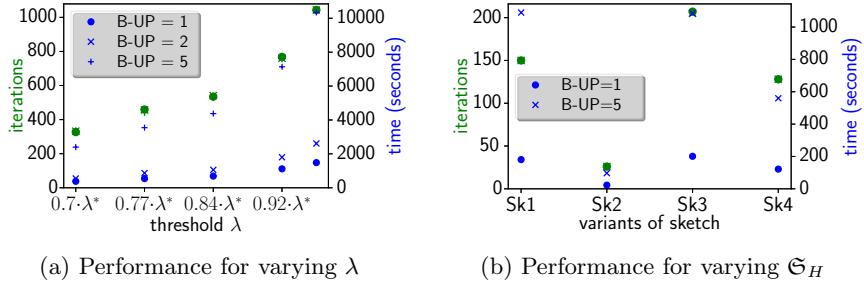
\begin{figure*}[t]
\centering
\subfloat[Performance for varying $\lambda$]{
\scalebox{.45}{
\input{figures/dpm_only_queue.pgf}
\label{fig:dpm2onlyqueue}
}
}	
\subfloat[Performance for varying $\Sketch$]{
\scalebox{.45}{
\input{figures/dpm_variants.pgf}
\label{fig:dpm2variants}
}
}
\vspace{-2mm}
\caption{Performance (runtime and iterations) on DPM}
\label{fig:scalability}
\vspace{-1mm}
\end{figure*}

\smallskip
\noindent\emph{DPM} has 9 holes with 260K realisations, and MCs have 5K (reachable) states on average, ranging from 2K to 8K~states. 
\emph{The performance of CEGIS significantly depends on the specification, namely, on the  thresholds appearing in the properties. } 
Fig.~\ref{fig:dpm2onlyqueue} shows how the number of iterations (left axis, green circle) and the runtime in seconds (right axis, blue) changes for varying $\lambda$ for property $\varphi_1$ (stars and crosses are explained later).
We obtain a speedup of $100{\times}$ over the baseline for $\lambda = 0.7{\cdot} \lambda^{*}$, dropping to $23{\times}$ for $\lambda=0.95{\cdot}\lambda^{*}$, where $\lambda^{*}$ is the minimal probabilty over all realisations.
The strong dependency between performance and ``unsatisfiability'' is not surprising.
The more unsatisfiable, the smaller the conflicts (as in~\cite{DBLP:conf/atva/DehnertJWAK14}). 
Small conflicts have a double beneficial effect.
First, the prototype uses an optimistic verifier searching for minimal conflicts; small conflicts are found faster than large ones.
Second, small conflicts prune more realisations. A slightly higher number of small conflicts yields a severe decrease in iterations.
Thus \emph{the further the threshold from the optimum, the better the performance}.

Reconsider Fig.~\ref{fig:dpm2onlyqueue}, crosses and stars correspond to a variant in which we have blown up the state space of the underlying MCs by a factor B-UP.
Observe that performance degrades similarly for the baseline and our algorithm, which means that \emph{the speedup w.r.t.\ the baseline is not considerably affected by the size of the underlying MCs.} 
This observation holds for various models and specifications.

Varying the sketch tremendously affects performance, cf.\ Fig.~\ref{fig:dpm2variants} for the performance of variants of the original sketch with some hole substituted. 
The framework performs significantly better on sketches with holes that lie in local regions of the MC. Holes relating to states all-over the MC are harder to prune. 
Finally, our prototype generally performs better with specifications that have multiple (conflicting) properties: Some realisations can be effectively pruned by conflicts w.r.t.\ property $\varphi_1$, whereas other realisations are easily pruned by conflicts w.r.t., e.g., property $\varphi_2$.

\smallskip\noindent\textsl{Intrusion} has 26 holes and 6800K realisations, the underlying MCs have only 500 states on average.
We observe an even more significant effect of the property thresholds on the performance, as the number of holes is larger (recall the optimistic verifier). We obtain a speedup of factor $2200$, $250$ and $5$ over the baseline, for thresholds $0.7{\cdot}\lambda^{*}$, $0.8{\cdot}\lambda^{*}$ and $0.9{\cdot}\lambda^{*}$, respectively. For $0.7{\cdot}\lambda^{*}$, many conflicts contain only 8 holes.
Blowing up the model does not affect the obtained speedups.
Differences among variants are again significant, albeit less extreme. 

\smallskip\noindent\textsl{Grid} is structurally different: only 6 holes in 3 commands and 1800 realisations, but MCs having 100K states on average. 
Observe that reaching the targets on expectation below some threshold implies that the goal must almost surely be reached. 
The MCs' topology and the few commands make pruning hard: 
 our algorithm needs more than $400$ iterations. 
 Still, we obtain a $3{\times}$ speedup for $\lambda = 0.98 {\cdot} \lambda^{*}$. 
 Pruning mostly follows from reasoning about realisations that do not reach the target almost surely. Therefore, the speedup is mostly independent of the relation between $\lambda$ and $\lambda^{*}$.

\medskip\noindent\textbf{Discussion.}
\emph{Optimistic verifiers} search for a minimal CE and thus solve an NP-hard problem~\cite{DBLP:journals/tocl/ChadhaV10,DBLP:journals/corr/abs-1305-5055}. In particular, we observed a lot of overhead when the smallest conflict is large, and any small CE that can be cheaply computed might be better for the performance (much like the computation of unsatisfiable cores in SMT solvers). Likewise, reusing information about holes from previous runs might benefit the performance.
Improvements in concise sketching, and exploiting the additional structure, will also improve performance.

\smallskip\noindent\emph{Sketching.} Families are simpler objects than sketches, but their explicit usage of states make them inadequate for modelling.
Families can be lifted to a (restricted) sketching class, as in~\cite{DBLP:conf/tacas/CeskaJJK19}. 
However, additional features like conflicts significantly ease the modelling process.
Consider \textsl{intrusion}: Without constraints, the number of realisations grows to $6{\cdot}10^{11}$ realisations. Put differently, the constraint allows to discard over $99.99\%$ of the realisations up front.
Moreover, constraints can exclude realisations that would yield unsupported programs, e.g, programs with infinite state spaces.
While modelling concise sketches with small underlying MCs, it may be hard to avoid such invalid realisations without the use of constraints.

\smallskip\noindent\emph{Comparison with CEGAR.}
We also compared with our \textsf{CEGAR-prototype}~\cite{DBLP:conf/tacas/CeskaJJK19}, which applies an abstraction-refinement loop towards, e.g., feasibility synthesis.
In particular, the abstraction aggregates multiple realisations in a single model to effectively  reason about a set of designs. This approach does not support multiple objective specification (as of now), and, more importantly, the algorithm is conceptually not capable of handling constraints. 
For \textsl{DPM} with a single objective, \textsf{CEGAR-prototype} is drastically faster, while on \textsl{Grid}, it is typically (often significantly)  slower.
Comparing two prototypes is intricate, but there is a CEGIS strength and weakness that  explains the different characteristics.  

\emph{Weakness:} 
Upon invocation, the CEGIS verifier gets exactly one realisation, and is (as of now)  \emph{unaware} of other options for the holes.
The verifier constructs a CE which is valid for all possible extensions (cf. Definition~\ref{def:genCE}), even for extensions which do not correspond to any realisation. It would be better if we compute CEs that are (only) valid for all possible realisations.
The following example (exaggerating \textsl{DPM}) illustrates that considering multiple realisations at once may be helpful:
Consider a family with a parametric transition (hole) from the initial state and specification that requires reaching the failure state with probability smaller than 0.1.
Assume that all 100 options lead to a failure states with probability 1. 
CEGIS never prunes this hole as it is relevant for every realisation. 
Knowing that all options lead to the failure state, however makes the hole trivially not relevant. Thus, all corresponding options can be safely  pruned.

\emph{Strength:} The weakness is related to its strength: the verifier  works with \emph{one} concrete MC.
An extreme example (exaggerating \textsl{Grid}) is a sketch with holes $h_0,\hdots, h_m$, where hole $h_0$ has options $1,\hdots, m$, and option $i$ makes holes $h_j$ $j \neq i$ irrelevant (by the model topology). 
CEGIS considers a realisation, say $\{ h_0 \mapsto i, \hdots, h_i \mapsto x, \hdots \}$, that violates the specification. As holes $h_j$ with $j \neq i$ are not relevant, CEGIS finds a conflict $\{ h_0 \mapsto i, h_i \mapsto x \}$.
Indeed, for every realisation, it is able to prune all but two holes. 
However, if the verifier would consider many realisations for $h_0$, it may (without advanced reasoning) generate much larger conflicts.
Thus, considering a single realisation naturally fixes the context of the  selected options and makes it clearer which holes are not relevant.

%% file: figures/dpm_only_queue.pgf
\begingroup%
\makeatletter%
\begin{pgfpicture}%
\pgfpathrectangle{\pgfpointorigin}{\pgfqpoint{5.000000in}{3.000000in}}%
\pgfusepath{use as bounding box, clip}%
\begin{pgfscope}%
\pgfsetbuttcap%
\pgfsetmiterjoin%
\definecolor{currentfill}{rgb}{1.000000,1.000000,1.000000}%
\pgfsetfillcolor{currentfill}%
\pgfsetlinewidth{0.000000pt}%
\definecolor{currentstroke}{rgb}{1.000000,1.000000,1.000000}%
\pgfsetstrokecolor{currentstroke}%
\pgfsetdash{}{0pt}%
\pgfpathmoveto{\pgfqpoint{0.000000in}{0.000000in}}%
\pgfpathlineto{\pgfqpoint{5.000000in}{0.000000in}}%
\pgfpathlineto{\pgfqpoint{5.000000in}{3.000000in}}%
\pgfpathlineto{\pgfqpoint{0.000000in}{3.000000in}}%
\pgfpathclose%
\pgfusepath{fill}%
\end{pgfscope}%
\begin{pgfscope}%
\pgfsetbuttcap%
\pgfsetmiterjoin%
\definecolor{currentfill}{rgb}{1.000000,1.000000,1.000000}%
\pgfsetfillcolor{currentfill}%
\pgfsetlinewidth{0.000000pt}%
\definecolor{currentstroke}{rgb}{0.000000,0.000000,0.000000}%
\pgfsetstrokecolor{currentstroke}%
\pgfsetstrokeopacity{0.000000}%
\pgfsetdash{}{0pt}%
\pgfpathmoveto{\pgfqpoint{0.907778in}{0.644809in}}%
\pgfpathlineto{\pgfqpoint{3.985972in}{0.644809in}}%
\pgfpathlineto{\pgfqpoint{3.985972in}{2.785000in}}%
\pgfpathlineto{\pgfqpoint{0.907778in}{2.785000in}}%
\pgfpathclose%
\pgfusepath{fill}%
\end{pgfscope}%
\begin{pgfscope}%
\pgfsetbuttcap%
\pgfsetroundjoin%
\definecolor{currentfill}{rgb}{0.000000,0.000000,0.000000}%
\pgfsetfillcolor{currentfill}%
\pgfsetlinewidth{0.803000pt}%
\definecolor{currentstroke}{rgb}{0.000000,0.000000,0.000000}%
\pgfsetstrokecolor{currentstroke}%
\pgfsetdash{}{0pt}%
\pgfsys@defobject{currentmarker}{\pgfqpoint{0.000000in}{-0.048611in}}{\pgfqpoint{0.000000in}{0.000000in}}{%
\pgfpathmoveto{\pgfqpoint{0.000000in}{0.000000in}}%
\pgfpathlineto{\pgfqpoint{0.000000in}{-0.048611in}}%
\pgfusepath{stroke,fill}%
}%
\begin{pgfscope}%
\pgfsys@transformshift{1.047696in}{0.644809in}%
\pgfsys@useobject{currentmarker}{}%
\end{pgfscope}%
\end{pgfscope}%
\begin{pgfscope}%
\pgftext[x=1.047696in,y=0.547587in,,top]{\sffamily\fontsize{16.000000}{14.400000}\selectfont $0.7 {\cdot} \lambda^{*}$}%
\end{pgfscope}%
\begin{pgfscope}%
\pgfsetbuttcap%
\pgfsetroundjoin%
\definecolor{currentfill}{rgb}{0.000000,0.000000,0.000000}%
\pgfsetfillcolor{currentfill}%
\pgfsetlinewidth{0.803000pt}%
\definecolor{currentstroke}{rgb}{0.000000,0.000000,0.000000}%
\pgfsetstrokecolor{currentstroke}%
\pgfsetdash{}{0pt}%
\pgfsys@defobject{currentmarker}{\pgfqpoint{0.000000in}{-0.048611in}}{\pgfqpoint{0.000000in}{0.000000in}}{%
\pgfpathmoveto{\pgfqpoint{0.000000in}{0.000000in}}%
\pgfpathlineto{\pgfqpoint{0.000000in}{-0.048611in}}%
\pgfusepath{stroke,fill}%
}%
\begin{pgfscope}%
\pgfsys@transformshift{1.847227in}{0.644809in}%
\pgfsys@useobject{currentmarker}{}%
\end{pgfscope}%
\end{pgfscope}%
\begin{pgfscope}%
\pgftext[x=1.847227in,y=0.547587in,,top]{\sffamily\fontsize{16.000000}{14.400000}\selectfont $0.77{\cdot}\lambda^{*}$}%
\end{pgfscope}%
\begin{pgfscope}%
\pgfsetbuttcap%
\pgfsetroundjoin%
\definecolor{currentfill}{rgb}{0.000000,0.000000,0.000000}%
\pgfsetfillcolor{currentfill}%
\pgfsetlinewidth{0.803000pt}%
\definecolor{currentstroke}{rgb}{0.000000,0.000000,0.000000}%
\pgfsetstrokecolor{currentstroke}%
\pgfsetdash{}{0pt}%
\pgfsys@defobject{currentmarker}{\pgfqpoint{0.000000in}{-0.048611in}}{\pgfqpoint{0.000000in}{0.000000in}}{%
\pgfpathmoveto{\pgfqpoint{0.000000in}{0.000000in}}%
\pgfpathlineto{\pgfqpoint{0.000000in}{-0.048611in}}%
\pgfusepath{stroke,fill}%
}%
\begin{pgfscope}%
\pgfsys@transformshift{2.646758in}{0.644809in}%
\pgfsys@useobject{currentmarker}{}%
\end{pgfscope}%
\end{pgfscope}%
\begin{pgfscope}%
\pgftext[x=2.646758in,y=0.547587in,,top]{\sffamily\fontsize{16.000000}{14.400000}\selectfont $0.84 {\cdot} \lambda^{*}$}%
\end{pgfscope}%
\begin{pgfscope}%
\pgfsetbuttcap%
\pgfsetroundjoin%
\definecolor{currentfill}{rgb}{0.000000,0.000000,0.000000}%
\pgfsetfillcolor{currentfill}%
\pgfsetlinewidth{0.803000pt}%
\definecolor{currentstroke}{rgb}{0.000000,0.000000,0.000000}%
\pgfsetstrokecolor{currentstroke}%
\pgfsetdash{}{0pt}%
\pgfsys@defobject{currentmarker}{\pgfqpoint{0.000000in}{-0.048611in}}{\pgfqpoint{0.000000in}{0.000000in}}{%
\pgfpathmoveto{\pgfqpoint{0.000000in}{0.000000in}}%
\pgfpathlineto{\pgfqpoint{0.000000in}{-0.048611in}}%
\pgfusepath{stroke,fill}%
}%
\begin{pgfscope}%
\pgfsys@transformshift{3.446289in}{0.644809in}%
\pgfsys@useobject{currentmarker}{}%
\end{pgfscope}%
\end{pgfscope}%
\begin{pgfscope}%
\pgftext[x=3.446289in,y=0.547587in,,top]{\sffamily\fontsize{16.000000}{14.400000}\selectfont $0.92{\cdot}\lambda^{*}$}%
\end{pgfscope}%
\begin{pgfscope}%
\pgftext[x=2.446875in,y=0.304031in,,top]{\sffamily\fontsize{16.000000}{14.400000}\selectfont threshold $\lambda$}%
\end{pgfscope}%
\begin{pgfscope}%
\pgfsetbuttcap%
\pgfsetroundjoin%
\definecolor{currentfill}{rgb}{0.000000,0.000000,0.000000}%
\pgfsetfillcolor{currentfill}%
\pgfsetlinewidth{0.803000pt}%
\definecolor{currentstroke}{rgb}{0.000000,0.000000,0.000000}%
\pgfsetstrokecolor{currentstroke}%
\pgfsetdash{}{0pt}%
\pgfsys@defobject{currentmarker}{\pgfqpoint{-0.048611in}{0.000000in}}{\pgfqpoint{0.000000in}{0.000000in}}{%
\pgfpathmoveto{\pgfqpoint{0.000000in}{0.000000in}}%
\pgfpathlineto{\pgfqpoint{-0.048611in}{0.000000in}}%
\pgfusepath{stroke,fill}%
}%
\begin{pgfscope}%
\pgfsys@transformshift{0.907778in}{0.644809in}%
\pgfsys@useobject{currentmarker}{}%
\end{pgfscope}%
\end{pgfscope}%
\begin{pgfscope}%
\pgftext[x=0.728889in,y=0.586976in,left,base]{\sffamily\fontsize{16.000000}{14.400000}\selectfont 0}%
\end{pgfscope}%
\begin{pgfscope}%
\pgfsetbuttcap%
\pgfsetroundjoin%
\definecolor{currentfill}{rgb}{0.000000,0.000000,0.000000}%
\pgfsetfillcolor{currentfill}%
\pgfsetlinewidth{0.803000pt}%
\definecolor{currentstroke}{rgb}{0.000000,0.000000,0.000000}%
\pgfsetstrokecolor{currentstroke}%
\pgfsetdash{}{0pt}%
\pgfsys@defobject{currentmarker}{\pgfqpoint{-0.048611in}{0.000000in}}{\pgfqpoint{0.000000in}{0.000000in}}{%
\pgfpathmoveto{\pgfqpoint{0.000000in}{0.000000in}}%
\pgfpathlineto{\pgfqpoint{-0.048611in}{0.000000in}}%
\pgfusepath{stroke,fill}%
}%
\begin{pgfscope}%
\pgfsys@transformshift{0.907778in}{1.041582in}%
\pgfsys@useobject{currentmarker}{}%
\end{pgfscope}%
\end{pgfscope}%
\begin{pgfscope}%
\pgftext[x=0.485556in,y=0.983748in,left,base]{\sffamily\fontsize{16.000000}{14.400000}\selectfont 200}%
\end{pgfscope}%
\begin{pgfscope}%
\pgfsetbuttcap%
\pgfsetroundjoin%
\definecolor{currentfill}{rgb}{0.000000,0.000000,0.000000}%
\pgfsetfillcolor{currentfill}%
\pgfsetlinewidth{0.803000pt}%
\definecolor{currentstroke}{rgb}{0.000000,0.000000,0.000000}%
\pgfsetstrokecolor{currentstroke}%
\pgfsetdash{}{0pt}%
\pgfsys@defobject{currentmarker}{\pgfqpoint{-0.048611in}{0.000000in}}{\pgfqpoint{0.000000in}{0.000000in}}{%
\pgfpathmoveto{\pgfqpoint{0.000000in}{0.000000in}}%
\pgfpathlineto{\pgfqpoint{-0.048611in}{0.000000in}}%
\pgfusepath{stroke,fill}%
}%
\begin{pgfscope}%
\pgfsys@transformshift{0.907778in}{1.438354in}%
\pgfsys@useobject{currentmarker}{}%
\end{pgfscope}%
\end{pgfscope}%
\begin{pgfscope}%
\pgftext[x=0.485556in,y=1.380521in,left,base]{\sffamily\fontsize{16.000000}{14.400000}\selectfont 400}%
\end{pgfscope}%
\begin{pgfscope}%
\pgfsetbuttcap%
\pgfsetroundjoin%
\definecolor{currentfill}{rgb}{0.000000,0.000000,0.000000}%
\pgfsetfillcolor{currentfill}%
\pgfsetlinewidth{0.803000pt}%
\definecolor{currentstroke}{rgb}{0.000000,0.000000,0.000000}%
\pgfsetstrokecolor{currentstroke}%
\pgfsetdash{}{0pt}%
\pgfsys@defobject{currentmarker}{\pgfqpoint{-0.048611in}{0.000000in}}{\pgfqpoint{0.000000in}{0.000000in}}{%
\pgfpathmoveto{\pgfqpoint{0.000000in}{0.000000in}}%
\pgfpathlineto{\pgfqpoint{-0.048611in}{0.000000in}}%
\pgfusepath{stroke,fill}%
}%
\begin{pgfscope}%
\pgfsys@transformshift{0.907778in}{1.835127in}%
\pgfsys@useobject{currentmarker}{}%
\end{pgfscope}%
\end{pgfscope}%
\begin{pgfscope}%
\pgftext[x=0.485556in,y=1.777293in,left,base]{\sffamily\fontsize{16.000000}{14.400000}\selectfont 600}%
\end{pgfscope}%
\begin{pgfscope}%
\pgfsetbuttcap%
\pgfsetroundjoin%
\definecolor{currentfill}{rgb}{0.000000,0.000000,0.000000}%
\pgfsetfillcolor{currentfill}%
\pgfsetlinewidth{0.803000pt}%
\definecolor{currentstroke}{rgb}{0.000000,0.000000,0.000000}%
\pgfsetstrokecolor{currentstroke}%
\pgfsetdash{}{0pt}%
\pgfsys@defobject{currentmarker}{\pgfqpoint{-0.048611in}{0.000000in}}{\pgfqpoint{0.000000in}{0.000000in}}{%
\pgfpathmoveto{\pgfqpoint{0.000000in}{0.000000in}}%
\pgfpathlineto{\pgfqpoint{-0.048611in}{0.000000in}}%
\pgfusepath{stroke,fill}%
}%
\begin{pgfscope}%
\pgfsys@transformshift{0.907778in}{2.231899in}%
\pgfsys@useobject{currentmarker}{}%
\end{pgfscope}%
\end{pgfscope}%
\begin{pgfscope}%
\pgftext[x=0.485556in,y=2.174066in,left,base]{\sffamily\fontsize{16.000000}{14.400000}\selectfont 800}%
\end{pgfscope}%
\begin{pgfscope}%
\pgfsetbuttcap%
\pgfsetroundjoin%
\definecolor{currentfill}{rgb}{0.000000,0.000000,0.000000}%
\pgfsetfillcolor{currentfill}%
\pgfsetlinewidth{0.803000pt}%
\definecolor{currentstroke}{rgb}{0.000000,0.000000,0.000000}%
\pgfsetstrokecolor{currentstroke}%
\pgfsetdash{}{0pt}%
\pgfsys@defobject{currentmarker}{\pgfqpoint{-0.048611in}{0.000000in}}{\pgfqpoint{0.000000in}{0.000000in}}{%
\pgfpathmoveto{\pgfqpoint{0.000000in}{0.000000in}}%
\pgfpathlineto{\pgfqpoint{-0.048611in}{0.000000in}}%
\pgfusepath{stroke,fill}%
}%
\begin{pgfscope}%
\pgfsys@transformshift{0.907778in}{2.628672in}%
\pgfsys@useobject{currentmarker}{}%
\end{pgfscope}%
\end{pgfscope}%
\begin{pgfscope}%
\pgftext[x=0.383889in,y=2.570838in,left,base]{\sffamily\fontsize{16.000000}{14.400000}\selectfont 1000}%
\end{pgfscope}%
\begin{pgfscope}%
\definecolor{textcolor}{rgb}{0.000000,0.500000,0.000000}%
\pgfsetstrokecolor{textcolor}%
\pgfsetfillcolor{textcolor}%
\pgftext[x=0.428333in,y=1.714905in,,bottom,rotate=90.000000]{\color{textcolor}\sffamily\fontsize{16.000000}{14.400000}\selectfont iterations}%
\end{pgfscope}%
\begin{pgfscope}%
\pgfpathrectangle{\pgfqpoint{0.907778in}{0.644809in}}{\pgfqpoint{3.078194in}{2.140191in}} %
\pgfusepath{clip}%
\pgfsetbuttcap%
\pgfsetroundjoin%
\definecolor{currentfill}{rgb}{0.000000,0.500000,0.000000}%
\pgfsetfillcolor{currentfill}%
\pgfsetlinewidth{1.003750pt}%
\definecolor{currentstroke}{rgb}{0.000000,0.500000,0.000000}%
\pgfsetstrokecolor{currentstroke}%
\pgfsetdash{}{0pt}%
\pgfsys@defobject{currentmarker}{\pgfqpoint{-0.041667in}{-0.041667in}}{\pgfqpoint{0.041667in}{0.041667in}}{%
\pgfpathmoveto{\pgfqpoint{0.000000in}{-0.041667in}}%
\pgfpathcurveto{\pgfqpoint{0.011050in}{-0.041667in}}{\pgfqpoint{0.021649in}{-0.037276in}}{\pgfqpoint{0.029463in}{-0.029463in}}%
\pgfpathcurveto{\pgfqpoint{0.037276in}{-0.021649in}}{\pgfqpoint{0.041667in}{-0.011050in}}{\pgfqpoint{0.041667in}{0.000000in}}%
\pgfpathcurveto{\pgfqpoint{0.041667in}{0.011050in}}{\pgfqpoint{0.037276in}{0.021649in}}{\pgfqpoint{0.029463in}{0.029463in}}%
\pgfpathcurveto{\pgfqpoint{0.021649in}{0.037276in}}{\pgfqpoint{0.011050in}{0.041667in}}{\pgfqpoint{0.000000in}{0.041667in}}%
\pgfpathcurveto{\pgfqpoint{-0.011050in}{0.041667in}}{\pgfqpoint{-0.021649in}{0.037276in}}{\pgfqpoint{-0.029463in}{0.029463in}}%
\pgfpathcurveto{\pgfqpoint{-0.037276in}{0.021649in}}{\pgfqpoint{-0.041667in}{0.011050in}}{\pgfqpoint{-0.041667in}{0.000000in}}%
\pgfpathcurveto{\pgfqpoint{-0.041667in}{-0.011050in}}{\pgfqpoint{-0.037276in}{-0.021649in}}{\pgfqpoint{-0.029463in}{-0.029463in}}%
\pgfpathcurveto{\pgfqpoint{-0.021649in}{-0.037276in}}{\pgfqpoint{-0.011050in}{-0.041667in}}{\pgfqpoint{0.000000in}{-0.041667in}}%
\pgfpathclose%
\pgfusepath{stroke,fill}%
}%
\begin{pgfscope}%
\pgfsys@transformshift{1.047696in}{1.293532in}%
\pgfsys@useobject{currentmarker}{}%
\end{pgfscope}%
\begin{pgfscope}%
\pgfsys@transformshift{1.847227in}{1.555402in}%
\pgfsys@useobject{currentmarker}{}%
\end{pgfscope}%
\begin{pgfscope}%
\pgfsys@transformshift{2.646758in}{1.706176in}%
\pgfsys@useobject{currentmarker}{}%
\end{pgfscope}%
\begin{pgfscope}%
\pgfsys@transformshift{3.446289in}{2.168416in}%
\pgfsys@useobject{currentmarker}{}%
\end{pgfscope}%
\begin{pgfscope}%
\pgfsys@transformshift{3.846054in}{2.713978in}%
\pgfsys@useobject{currentmarker}{}%
\end{pgfscope}%
\end{pgfscope}%
\begin{pgfscope}%
\pgfpathrectangle{\pgfqpoint{0.907778in}{0.644809in}}{\pgfqpoint{3.078194in}{2.140191in}} %
\pgfusepath{clip}%
\pgfsetbuttcap%
\pgfsetroundjoin%
\definecolor{currentfill}{rgb}{0.000000,0.500000,0.000000}%
\pgfsetfillcolor{currentfill}%
\pgfsetlinewidth{1.003750pt}%
\definecolor{currentstroke}{rgb}{0.000000,0.500000,0.000000}%
\pgfsetstrokecolor{currentstroke}%
\pgfsetdash{}{0pt}%
\pgfsys@defobject{currentmarker}{\pgfqpoint{-0.041667in}{-0.041667in}}{\pgfqpoint{0.041667in}{0.041667in}}{%
\pgfpathmoveto{\pgfqpoint{-0.041667in}{-0.041667in}}%
\pgfpathlineto{\pgfqpoint{0.041667in}{0.041667in}}%
\pgfpathmoveto{\pgfqpoint{-0.041667in}{0.041667in}}%
\pgfpathlineto{\pgfqpoint{0.041667in}{-0.041667in}}%
\pgfusepath{stroke,fill}%
}%
\begin{pgfscope}%
\pgfsys@transformshift{1.047696in}{1.303451in}%
\pgfsys@useobject{currentmarker}{}%
\end{pgfscope}%
\begin{pgfscope}%
\pgfsys@transformshift{1.847227in}{1.551434in}%
\pgfsys@useobject{currentmarker}{}%
\end{pgfscope}%
\begin{pgfscope}%
\pgfsys@transformshift{2.646758in}{1.718079in}%
\pgfsys@useobject{currentmarker}{}%
\end{pgfscope}%
\begin{pgfscope}%
\pgfsys@transformshift{3.446289in}{2.152545in}%
\pgfsys@useobject{currentmarker}{}%
\end{pgfscope}%
\begin{pgfscope}%
\pgfsys@transformshift{3.846054in}{2.713978in}%
\pgfsys@useobject{currentmarker}{}%
\end{pgfscope}%
\end{pgfscope}%
\begin{pgfscope}%
\pgfpathrectangle{\pgfqpoint{0.907778in}{0.644809in}}{\pgfqpoint{3.078194in}{2.140191in}} %
\pgfusepath{clip}%
\pgfsetbuttcap%
\pgfsetroundjoin%
\definecolor{currentfill}{rgb}{0.000000,0.500000,0.000000}%
\pgfsetfillcolor{currentfill}%
\pgfsetlinewidth{1.003750pt}%
\definecolor{currentstroke}{rgb}{0.000000,0.500000,0.000000}%
\pgfsetstrokecolor{currentstroke}%
\pgfsetdash{}{0pt}%
\pgfsys@defobject{currentmarker}{\pgfqpoint{-0.041667in}{-0.041667in}}{\pgfqpoint{0.041667in}{0.041667in}}{%
\pgfpathmoveto{\pgfqpoint{-0.041667in}{0.000000in}}%
\pgfpathlineto{\pgfqpoint{0.041667in}{0.000000in}}%
\pgfpathmoveto{\pgfqpoint{0.000000in}{-0.041667in}}%
\pgfpathlineto{\pgfqpoint{0.000000in}{0.041667in}}%
\pgfusepath{stroke,fill}%
}%
\begin{pgfscope}%
\pgfsys@transformshift{1.047696in}{1.307419in}%
\pgfsys@useobject{currentmarker}{}%
\end{pgfscope}%
\begin{pgfscope}%
\pgfsys@transformshift{1.847227in}{1.523660in}%
\pgfsys@useobject{currentmarker}{}%
\end{pgfscope}%
\begin{pgfscope}%
\pgfsys@transformshift{2.646758in}{1.696256in}%
\pgfsys@useobject{currentmarker}{}%
\end{pgfscope}%
\begin{pgfscope}%
\pgfsys@transformshift{3.446289in}{2.148577in}%
\pgfsys@useobject{currentmarker}{}%
\end{pgfscope}%
\begin{pgfscope}%
\pgfsys@transformshift{3.846054in}{2.713978in}%
\pgfsys@useobject{currentmarker}{}%
\end{pgfscope}%
\end{pgfscope}%
\begin{pgfscope}%
\pgfsetrectcap%
\pgfsetmiterjoin%
\pgfsetlinewidth{0.803000pt}%
\definecolor{currentstroke}{rgb}{0.000000,0.000000,0.000000}%
\pgfsetstrokecolor{currentstroke}%
\pgfsetdash{}{0pt}%
\pgfpathmoveto{\pgfqpoint{0.907778in}{0.644809in}}%
\pgfpathlineto{\pgfqpoint{0.907778in}{2.785000in}}%
\pgfusepath{stroke}%
\end{pgfscope}%
\begin{pgfscope}%
\pgfsetrectcap%
\pgfsetmiterjoin%
\pgfsetlinewidth{0.803000pt}%
\definecolor{currentstroke}{rgb}{0.000000,0.000000,0.000000}%
\pgfsetstrokecolor{currentstroke}%
\pgfsetdash{}{0pt}%
\pgfpathmoveto{\pgfqpoint{3.985972in}{0.644809in}}%
\pgfpathlineto{\pgfqpoint{3.985972in}{2.785000in}}%
\pgfusepath{stroke}%
\end{pgfscope}%
\begin{pgfscope}%
\pgfsetrectcap%
\pgfsetmiterjoin%
\pgfsetlinewidth{0.803000pt}%
\definecolor{currentstroke}{rgb}{0.000000,0.000000,0.000000}%
\pgfsetstrokecolor{currentstroke}%
\pgfsetdash{}{0pt}%
\pgfpathmoveto{\pgfqpoint{0.907778in}{0.644809in}}%
\pgfpathlineto{\pgfqpoint{3.985972in}{0.644809in}}%
\pgfusepath{stroke}%
\end{pgfscope}%
\begin{pgfscope}%
\pgfsetrectcap%
\pgfsetmiterjoin%
\pgfsetlinewidth{0.803000pt}%
\definecolor{currentstroke}{rgb}{0.000000,0.000000,0.000000}%
\pgfsetstrokecolor{currentstroke}%
\pgfsetdash{}{0pt}%
\pgfpathmoveto{\pgfqpoint{0.907778in}{2.785000in}}%
\pgfpathlineto{\pgfqpoint{3.985972in}{2.785000in}}%
\pgfusepath{stroke}%
\end{pgfscope}%
\begin{pgfscope}%
\pgfsetbuttcap%
\pgfsetroundjoin%
\definecolor{currentfill}{rgb}{0.000000,0.000000,0.000000}%
\pgfsetfillcolor{currentfill}%
\pgfsetlinewidth{0.803000pt}%
\definecolor{currentstroke}{rgb}{0.000000,0.000000,0.000000}%
\pgfsetstrokecolor{currentstroke}%
\pgfsetdash{}{0pt}%
\pgfsys@defobject{currentmarker}{\pgfqpoint{0.000000in}{0.000000in}}{\pgfqpoint{0.048611in}{0.000000in}}{%
\pgfpathmoveto{\pgfqpoint{0.000000in}{0.000000in}}%
\pgfpathlineto{\pgfqpoint{0.048611in}{0.000000in}}%
\pgfusepath{stroke,fill}%
}%
\begin{pgfscope}%
\pgfsys@transformshift{3.985972in}{0.644809in}%
\pgfsys@useobject{currentmarker}{}%
\end{pgfscope}%
\end{pgfscope}%
\begin{pgfscope}%
\pgftext[x=4.083194in,y=0.586976in,left,base]{\sffamily\fontsize{16.000000}{14.400000}\selectfont 0}%
\end{pgfscope}%
\begin{pgfscope}%
\pgfsetbuttcap%
\pgfsetroundjoin%
\definecolor{currentfill}{rgb}{0.000000,0.000000,0.000000}%
\pgfsetfillcolor{currentfill}%
\pgfsetlinewidth{0.803000pt}%
\definecolor{currentstroke}{rgb}{0.000000,0.000000,0.000000}%
\pgfsetstrokecolor{currentstroke}%
\pgfsetdash{}{0pt}%
\pgfsys@defobject{currentmarker}{\pgfqpoint{0.000000in}{0.000000in}}{\pgfqpoint{0.048611in}{0.000000in}}{%
\pgfpathmoveto{\pgfqpoint{0.000000in}{0.000000in}}%
\pgfpathlineto{\pgfqpoint{0.048611in}{0.000000in}}%
\pgfusepath{stroke,fill}%
}%
\begin{pgfscope}%
\pgfsys@transformshift{3.985972in}{1.039483in}%
\pgfsys@useobject{currentmarker}{}%
\end{pgfscope}%
\end{pgfscope}%
\begin{pgfscope}%
\pgftext[x=4.083194in,y=0.981650in,left,base]{\sffamily\fontsize{16.000000}{14.400000}\selectfont 2000}%
\end{pgfscope}%
\begin{pgfscope}%
\pgfsetbuttcap%
\pgfsetroundjoin%
\definecolor{currentfill}{rgb}{0.000000,0.000000,0.000000}%
\pgfsetfillcolor{currentfill}%
\pgfsetlinewidth{0.803000pt}%
\definecolor{currentstroke}{rgb}{0.000000,0.000000,0.000000}%
\pgfsetstrokecolor{currentstroke}%
\pgfsetdash{}{0pt}%
\pgfsys@defobject{currentmarker}{\pgfqpoint{0.000000in}{0.000000in}}{\pgfqpoint{0.048611in}{0.000000in}}{%
\pgfpathmoveto{\pgfqpoint{0.000000in}{0.000000in}}%
\pgfpathlineto{\pgfqpoint{0.048611in}{0.000000in}}%
\pgfusepath{stroke,fill}%
}%
\begin{pgfscope}%
\pgfsys@transformshift{3.985972in}{1.434157in}%
\pgfsys@useobject{currentmarker}{}%
\end{pgfscope}%
\end{pgfscope}%
\begin{pgfscope}%
\pgftext[x=4.083194in,y=1.376324in,left,base]{\sffamily\fontsize{16.000000}{14.400000}\selectfont 4000}%
\end{pgfscope}%
\begin{pgfscope}%
\pgfsetbuttcap%
\pgfsetroundjoin%
\definecolor{currentfill}{rgb}{0.000000,0.000000,0.000000}%
\pgfsetfillcolor{currentfill}%
\pgfsetlinewidth{0.803000pt}%
\definecolor{currentstroke}{rgb}{0.000000,0.000000,0.000000}%
\pgfsetstrokecolor{currentstroke}%
\pgfsetdash{}{0pt}%
\pgfsys@defobject{currentmarker}{\pgfqpoint{0.000000in}{0.000000in}}{\pgfqpoint{0.048611in}{0.000000in}}{%
\pgfpathmoveto{\pgfqpoint{0.000000in}{0.000000in}}%
\pgfpathlineto{\pgfqpoint{0.048611in}{0.000000in}}%
\pgfusepath{stroke,fill}%
}%
\begin{pgfscope}%
\pgfsys@transformshift{3.985972in}{1.828832in}%
\pgfsys@useobject{currentmarker}{}%
\end{pgfscope}%
\end{pgfscope}%
\begin{pgfscope}%
\pgftext[x=4.083194in,y=1.770998in,left,base]{\sffamily\fontsize{16.000000}{14.400000}\selectfont 6000}%
\end{pgfscope}%
\begin{pgfscope}%
\pgfsetbuttcap%
\pgfsetroundjoin%
\definecolor{currentfill}{rgb}{0.000000,0.000000,0.000000}%
\pgfsetfillcolor{currentfill}%
\pgfsetlinewidth{0.803000pt}%
\definecolor{currentstroke}{rgb}{0.000000,0.000000,0.000000}%
\pgfsetstrokecolor{currentstroke}%
\pgfsetdash{}{0pt}%
\pgfsys@defobject{currentmarker}{\pgfqpoint{0.000000in}{0.000000in}}{\pgfqpoint{0.048611in}{0.000000in}}{%
\pgfpathmoveto{\pgfqpoint{0.000000in}{0.000000in}}%
\pgfpathlineto{\pgfqpoint{0.048611in}{0.000000in}}%
\pgfusepath{stroke,fill}%
}%
\begin{pgfscope}%
\pgfsys@transformshift{3.985972in}{2.223506in}%
\pgfsys@useobject{currentmarker}{}%
\end{pgfscope}%
\end{pgfscope}%
\begin{pgfscope}%
\pgftext[x=4.083194in,y=2.165673in,left,base]{\sffamily\fontsize{16.000000}{14.400000}\selectfont 8000}%
\end{pgfscope}%
\begin{pgfscope}%
\pgfsetbuttcap%
\pgfsetroundjoin%
\definecolor{currentfill}{rgb}{0.000000,0.000000,0.000000}%
\pgfsetfillcolor{currentfill}%
\pgfsetlinewidth{0.803000pt}%
\definecolor{currentstroke}{rgb}{0.000000,0.000000,0.000000}%
\pgfsetstrokecolor{currentstroke}%
\pgfsetdash{}{0pt}%
\pgfsys@defobject{currentmarker}{\pgfqpoint{0.000000in}{0.000000in}}{\pgfqpoint{0.048611in}{0.000000in}}{%
\pgfpathmoveto{\pgfqpoint{0.000000in}{0.000000in}}%
\pgfpathlineto{\pgfqpoint{0.048611in}{0.000000in}}%
\pgfusepath{stroke,fill}%
}%
\begin{pgfscope}%
\pgfsys@transformshift{3.985972in}{2.618180in}%
\pgfsys@useobject{currentmarker}{}%
\end{pgfscope}%
\end{pgfscope}%
\begin{pgfscope}%
\pgftext[x=4.083194in,y=2.560347in,left,base]{\sffamily\fontsize{16.000000}{14.400000}\selectfont 10000}%
\end{pgfscope}%
\begin{pgfscope}%
\definecolor{textcolor}{rgb}{0.000000,0.000000,1.000000}%
\pgfsetstrokecolor{textcolor}%
\pgfsetfillcolor{textcolor}%
\pgftext[x=4.647083in,y=1.714905in,,top,rotate=90.000000]{\color{textcolor}\sffamily\fontsize{16.000000}{14.400000}\selectfont time (seconds)}%
\end{pgfscope}%
\begin{pgfscope}%
\pgfpathrectangle{\pgfqpoint{0.907778in}{0.644809in}}{\pgfqpoint{3.078194in}{2.140191in}} %
\pgfusepath{clip}%
\pgfsetbuttcap%
\pgfsetroundjoin%
\definecolor{currentfill}{rgb}{0.000000,0.000000,1.000000}%
\pgfsetfillcolor{currentfill}%
\pgfsetlinewidth{1.003750pt}%
\definecolor{currentstroke}{rgb}{0.000000,0.000000,1.000000}%
\pgfsetstrokecolor{currentstroke}%
\pgfsetdash{}{0pt}%
\pgfsys@defobject{currentmarker}{\pgfqpoint{-0.034722in}{-0.034722in}}{\pgfqpoint{0.034722in}{0.034722in}}{%
\pgfpathmoveto{\pgfqpoint{0.000000in}{-0.034722in}}%
\pgfpathcurveto{\pgfqpoint{0.009208in}{-0.034722in}}{\pgfqpoint{0.018041in}{-0.031064in}}{\pgfqpoint{0.024552in}{-0.024552in}}%
\pgfpathcurveto{\pgfqpoint{0.031064in}{-0.018041in}}{\pgfqpoint{0.034722in}{-0.009208in}}{\pgfqpoint{0.034722in}{0.000000in}}%
\pgfpathcurveto{\pgfqpoint{0.034722in}{0.009208in}}{\pgfqpoint{0.031064in}{0.018041in}}{\pgfqpoint{0.024552in}{0.024552in}}%
\pgfpathcurveto{\pgfqpoint{0.018041in}{0.031064in}}{\pgfqpoint{0.009208in}{0.034722in}}{\pgfqpoint{0.000000in}{0.034722in}}%
\pgfpathcurveto{\pgfqpoint{-0.009208in}{0.034722in}}{\pgfqpoint{-0.018041in}{0.031064in}}{\pgfqpoint{-0.024552in}{0.024552in}}%
\pgfpathcurveto{\pgfqpoint{-0.031064in}{0.018041in}}{\pgfqpoint{-0.034722in}{0.009208in}}{\pgfqpoint{-0.034722in}{0.000000in}}%
\pgfpathcurveto{\pgfqpoint{-0.034722in}{-0.009208in}}{\pgfqpoint{-0.031064in}{-0.018041in}}{\pgfqpoint{-0.024552in}{-0.024552in}}%
\pgfpathcurveto{\pgfqpoint{-0.018041in}{-0.031064in}}{\pgfqpoint{-0.009208in}{-0.034722in}}{\pgfqpoint{0.000000in}{-0.034722in}}%
\pgfpathclose%
\pgfusepath{stroke,fill}%
}%
\begin{pgfscope}%
\pgfsys@transformshift{1.047696in}{0.720081in}%
\pgfsys@useobject{currentmarker}{}%
\end{pgfscope}%
\begin{pgfscope}%
\pgfsys@transformshift{1.847227in}{0.751848in}%
\pgfsys@useobject{currentmarker}{}%
\end{pgfscope}%
\begin{pgfscope}%
\pgfsys@transformshift{2.646758in}{0.780885in}%
\pgfsys@useobject{currentmarker}{}%
\end{pgfscope}%
\begin{pgfscope}%
\pgfsys@transformshift{3.446289in}{0.865275in}%
\pgfsys@useobject{currentmarker}{}%
\end{pgfscope}%
\begin{pgfscope}%
\pgfsys@transformshift{3.846054in}{0.937775in}%
\pgfsys@useobject{currentmarker}{}%
\end{pgfscope}%
\end{pgfscope}%
\begin{pgfscope}%
\pgfpathrectangle{\pgfqpoint{0.907778in}{0.644809in}}{\pgfqpoint{3.078194in}{2.140191in}} %
\pgfusepath{clip}%
\pgfsetbuttcap%
\pgfsetroundjoin%
\definecolor{currentfill}{rgb}{0.000000,0.000000,1.000000}%
\pgfsetfillcolor{currentfill}%
\pgfsetlinewidth{1.003750pt}%
\definecolor{currentstroke}{rgb}{0.000000,0.000000,1.000000}%
\pgfsetstrokecolor{currentstroke}%
\pgfsetdash{}{0pt}%
\pgfsys@defobject{currentmarker}{\pgfqpoint{-0.034722in}{-0.034722in}}{\pgfqpoint{0.034722in}{0.034722in}}{%
\pgfpathmoveto{\pgfqpoint{-0.034722in}{-0.034722in}}%
\pgfpathlineto{\pgfqpoint{0.034722in}{0.034722in}}%
\pgfpathmoveto{\pgfqpoint{-0.034722in}{0.034722in}}%
\pgfpathlineto{\pgfqpoint{0.034722in}{-0.034722in}}%
\pgfusepath{stroke,fill}%
}%
\begin{pgfscope}%
\pgfsys@transformshift{1.047696in}{0.753137in}%
\pgfsys@useobject{currentmarker}{}%
\end{pgfscope}%
\begin{pgfscope}%
\pgfsys@transformshift{1.847227in}{0.816507in}%
\pgfsys@useobject{currentmarker}{}%
\end{pgfscope}%
\begin{pgfscope}%
\pgfsys@transformshift{2.646758in}{0.852978in}%
\pgfsys@useobject{currentmarker}{}%
\end{pgfscope}%
\begin{pgfscope}%
\pgfsys@transformshift{3.446289in}{1.000793in}%
\pgfsys@useobject{currentmarker}{}%
\end{pgfscope}%
\begin{pgfscope}%
\pgfsys@transformshift{3.846054in}{1.160527in}%
\pgfsys@useobject{currentmarker}{}%
\end{pgfscope}%
\end{pgfscope}%
\begin{pgfscope}%
\pgfpathrectangle{\pgfqpoint{0.907778in}{0.644809in}}{\pgfqpoint{3.078194in}{2.140191in}} %
\pgfusepath{clip}%
\pgfsetbuttcap%
\pgfsetroundjoin%
\definecolor{currentfill}{rgb}{0.000000,0.000000,1.000000}%
\pgfsetfillcolor{currentfill}%
\pgfsetlinewidth{1.003750pt}%
\definecolor{currentstroke}{rgb}{0.000000,0.000000,1.000000}%
\pgfsetstrokecolor{currentstroke}%
\pgfsetdash{}{0pt}%
\pgfsys@defobject{currentmarker}{\pgfqpoint{-0.034722in}{-0.034722in}}{\pgfqpoint{0.034722in}{0.034722in}}{%
\pgfpathmoveto{\pgfqpoint{-0.034722in}{0.000000in}}%
\pgfpathlineto{\pgfqpoint{0.034722in}{0.000000in}}%
\pgfpathmoveto{\pgfqpoint{0.000000in}{-0.034722in}}%
\pgfpathlineto{\pgfqpoint{0.000000in}{0.034722in}}%
\pgfusepath{stroke,fill}%
}%
\begin{pgfscope}%
\pgfsys@transformshift{1.047696in}{1.118660in}%
\pgfsys@useobject{currentmarker}{}%
\end{pgfscope}%
\begin{pgfscope}%
\pgfsys@transformshift{1.847227in}{1.344478in}%
\pgfsys@useobject{currentmarker}{}%
\end{pgfscope}%
\begin{pgfscope}%
\pgfsys@transformshift{2.646758in}{1.508126in}%
\pgfsys@useobject{currentmarker}{}%
\end{pgfscope}%
\begin{pgfscope}%
\pgfsys@transformshift{3.446289in}{2.053022in}%
\pgfsys@useobject{currentmarker}{}%
\end{pgfscope}%
\begin{pgfscope}%
\pgfsys@transformshift{3.846054in}{2.686671in}%
\pgfsys@useobject{currentmarker}{}%
\end{pgfscope}%
\end{pgfscope}%
\begin{pgfscope}%
\pgfsetrectcap%
\pgfsetmiterjoin%
\pgfsetlinewidth{0.803000pt}%
\definecolor{currentstroke}{rgb}{0.000000,0.000000,0.000000}%
\pgfsetstrokecolor{currentstroke}%
\pgfsetdash{}{0pt}%
\pgfpathmoveto{\pgfqpoint{0.907778in}{0.644809in}}%
\pgfpathlineto{\pgfqpoint{0.907778in}{2.785000in}}%
\pgfusepath{stroke}%
\end{pgfscope}%
\begin{pgfscope}%
\pgfsetrectcap%
\pgfsetmiterjoin%
\pgfsetlinewidth{0.803000pt}%
\definecolor{currentstroke}{rgb}{0.000000,0.000000,0.000000}%
\pgfsetstrokecolor{currentstroke}%
\pgfsetdash{}{0pt}%
\pgfpathmoveto{\pgfqpoint{3.985972in}{0.644809in}}%
\pgfpathlineto{\pgfqpoint{3.985972in}{2.785000in}}%
\pgfusepath{stroke}%
\end{pgfscope}%
\begin{pgfscope}%
\pgfsetrectcap%
\pgfsetmiterjoin%
\pgfsetlinewidth{0.803000pt}%
\definecolor{currentstroke}{rgb}{0.000000,0.000000,0.000000}%
\pgfsetstrokecolor{currentstroke}%
\pgfsetdash{}{0pt}%
\pgfpathmoveto{\pgfqpoint{0.907778in}{0.644809in}}%
\pgfpathlineto{\pgfqpoint{3.985972in}{0.644809in}}%
\pgfusepath{stroke}%
\end{pgfscope}%
\begin{pgfscope}%
\pgfsetrectcap%
\pgfsetmiterjoin%
\pgfsetlinewidth{0.803000pt}%
\definecolor{currentstroke}{rgb}{0.000000,0.000000,0.000000}%
\pgfsetstrokecolor{currentstroke}%
\pgfsetdash{}{0pt}%
\pgfpathmoveto{\pgfqpoint{0.907778in}{2.785000in}}%
\pgfpathlineto{\pgfqpoint{3.985972in}{2.785000in}}%
\pgfusepath{stroke}%
\end{pgfscope}%
\begin{pgfscope}%
\pgfsetbuttcap%
\pgfsetmiterjoin%
\definecolor{currentfill}{rgb}{0.300000,0.300000,0.300000}%
\pgfsetfillcolor{currentfill}%
\pgfsetfillopacity{0.500000}%
\pgfsetlinewidth{1.003750pt}%
\definecolor{currentstroke}{rgb}{0.300000,0.300000,0.300000}%
\pgfsetstrokecolor{currentstroke}%
\pgfsetstrokeopacity{0.500000}%
\pgfsetdash{}{0pt}%
\pgfpathmoveto{\pgfqpoint{1.052222in}{1.926889in}}%
\pgfpathlineto{\pgfqpoint{2.464556in}{1.926889in}}%
\pgfpathquadraticcurveto{\pgfqpoint{2.497889in}{1.926889in}}{\pgfqpoint{2.497889in}{1.960223in}}%
\pgfpathlineto{\pgfqpoint{2.497889in}{2.640556in}}%
\pgfpathquadraticcurveto{\pgfqpoint{2.497889in}{2.673889in}}{\pgfqpoint{2.464556in}{2.673889in}}%
\pgfpathlineto{\pgfqpoint{1.052222in}{2.673889in}}%
\pgfpathquadraticcurveto{\pgfqpoint{1.018889in}{2.673889in}}{\pgfqpoint{1.018889in}{2.640556in}}%
\pgfpathlineto{\pgfqpoint{1.018889in}{1.960223in}}%
\pgfpathquadraticcurveto{\pgfqpoint{1.018889in}{1.926889in}}{\pgfqpoint{1.052222in}{1.926889in}}%
\pgfpathclose%
\pgfusepath{stroke,fill}%
\end{pgfscope}%
\begin{pgfscope}%
\pgfsetbuttcap%
\pgfsetmiterjoin%
\definecolor{currentfill}{rgb}{1.000000,1.000000,1.000000}%
\pgfsetfillcolor{currentfill}%
\pgfsetfillopacity{0.500000}%
\pgfsetlinewidth{1.003750pt}%
\definecolor{currentstroke}{rgb}{0.800000,0.800000,0.800000}%
\pgfsetstrokecolor{currentstroke}%
\pgfsetstrokeopacity{0.500000}%
\pgfsetdash{}{0pt}%
\pgfpathmoveto{\pgfqpoint{1.024444in}{1.954667in}}%
\pgfpathlineto{\pgfqpoint{2.436778in}{1.954667in}}%
\pgfpathquadraticcurveto{\pgfqpoint{2.470111in}{1.954667in}}{\pgfqpoint{2.470111in}{1.988000in}}%
\pgfpathlineto{\pgfqpoint{2.470111in}{2.668333in}}%
\pgfpathquadraticcurveto{\pgfqpoint{2.470111in}{2.701667in}}{\pgfqpoint{2.436778in}{2.701667in}}%
\pgfpathlineto{\pgfqpoint{1.024444in}{2.701667in}}%
\pgfpathquadraticcurveto{\pgfqpoint{0.991111in}{2.701667in}}{\pgfqpoint{0.991111in}{2.668333in}}%
\pgfpathlineto{\pgfqpoint{0.991111in}{1.988000in}}%
\pgfpathquadraticcurveto{\pgfqpoint{0.991111in}{1.954667in}}{\pgfqpoint{1.024444in}{1.954667in}}%
\pgfpathclose%
\pgfusepath{stroke,fill}%
\end{pgfscope}%
\begin{pgfscope}%
\pgfsetbuttcap%
\pgfsetroundjoin%
\definecolor{currentfill}{rgb}{0.000000,0.000000,1.000000}%
\pgfsetfillcolor{currentfill}%
\pgfsetlinewidth{1.003750pt}%
\definecolor{currentstroke}{rgb}{0.000000,0.000000,1.000000}%
\pgfsetstrokecolor{currentstroke}%
\pgfsetdash{}{0pt}%
\pgfsys@defobject{currentmarker}{\pgfqpoint{-0.034722in}{-0.034722in}}{\pgfqpoint{0.034722in}{0.034722in}}{%
\pgfpathmoveto{\pgfqpoint{0.000000in}{-0.034722in}}%
\pgfpathcurveto{\pgfqpoint{0.009208in}{-0.034722in}}{\pgfqpoint{0.018041in}{-0.031064in}}{\pgfqpoint{0.024552in}{-0.024552in}}%
\pgfpathcurveto{\pgfqpoint{0.031064in}{-0.018041in}}{\pgfqpoint{0.034722in}{-0.009208in}}{\pgfqpoint{0.034722in}{0.000000in}}%
\pgfpathcurveto{\pgfqpoint{0.034722in}{0.009208in}}{\pgfqpoint{0.031064in}{0.018041in}}{\pgfqpoint{0.024552in}{0.024552in}}%
\pgfpathcurveto{\pgfqpoint{0.018041in}{0.031064in}}{\pgfqpoint{0.009208in}{0.034722in}}{\pgfqpoint{0.000000in}{0.034722in}}%
\pgfpathcurveto{\pgfqpoint{-0.009208in}{0.034722in}}{\pgfqpoint{-0.018041in}{0.031064in}}{\pgfqpoint{-0.024552in}{0.024552in}}%
\pgfpathcurveto{\pgfqpoint{-0.031064in}{0.018041in}}{\pgfqpoint{-0.034722in}{0.009208in}}{\pgfqpoint{-0.034722in}{0.000000in}}%
\pgfpathcurveto{\pgfqpoint{-0.034722in}{-0.009208in}}{\pgfqpoint{-0.031064in}{-0.018041in}}{\pgfqpoint{-0.024552in}{-0.024552in}}%
\pgfpathcurveto{\pgfqpoint{-0.018041in}{-0.031064in}}{\pgfqpoint{-0.009208in}{-0.034722in}}{\pgfqpoint{0.000000in}{-0.034722in}}%
\pgfpathclose%
\pgfusepath{stroke,fill}%
}%
\begin{pgfscope}%
\pgfsys@transformshift{1.224444in}{2.576667in}%
\pgfsys@useobject{currentmarker}{}%
\end{pgfscope}%
\end{pgfscope}%
\begin{pgfscope}%
\pgftext[x=1.524444in,y=2.518333in,left,base]{\sffamily\fontsize{14.000000}{14.400000}\selectfont B-UP = 1}%
\end{pgfscope}%
\begin{pgfscope}%
\pgfsetbuttcap%
\pgfsetroundjoin%
\definecolor{currentfill}{rgb}{0.000000,0.000000,1.000000}%
\pgfsetfillcolor{currentfill}%
\pgfsetlinewidth{1.003750pt}%
\definecolor{currentstroke}{rgb}{0.000000,0.000000,1.000000}%
\pgfsetstrokecolor{currentstroke}%
\pgfsetdash{}{0pt}%
\pgfsys@defobject{currentmarker}{\pgfqpoint{-0.034722in}{-0.034722in}}{\pgfqpoint{0.034722in}{0.034722in}}{%
\pgfpathmoveto{\pgfqpoint{-0.034722in}{-0.034722in}}%
\pgfpathlineto{\pgfqpoint{0.034722in}{0.034722in}}%
\pgfpathmoveto{\pgfqpoint{-0.034722in}{0.034722in}}%
\pgfpathlineto{\pgfqpoint{0.034722in}{-0.034722in}}%
\pgfusepath{stroke,fill}%
}%
\begin{pgfscope}%
\pgfsys@transformshift{1.224444in}{2.344333in}%
\pgfsys@useobject{currentmarker}{}%
\end{pgfscope}%
\end{pgfscope}%
\begin{pgfscope}%
\pgftext[x=1.524444in,y=2.286000in,left,base]{\sffamily\fontsize{14.000000}{14.400000}\selectfont B-UP = 2}%
\end{pgfscope}%
\begin{pgfscope}%
\pgfsetbuttcap%
\pgfsetroundjoin%
\definecolor{currentfill}{rgb}{0.000000,0.000000,1.000000}%
\pgfsetfillcolor{currentfill}%
\pgfsetlinewidth{1.003750pt}%
\definecolor{currentstroke}{rgb}{0.000000,0.000000,1.000000}%
\pgfsetstrokecolor{currentstroke}%
\pgfsetdash{}{0pt}%
\pgfsys@defobject{currentmarker}{\pgfqpoint{-0.034722in}{-0.034722in}}{\pgfqpoint{0.034722in}{0.034722in}}{%
\pgfpathmoveto{\pgfqpoint{-0.034722in}{0.000000in}}%
\pgfpathlineto{\pgfqpoint{0.034722in}{0.000000in}}%
\pgfpathmoveto{\pgfqpoint{0.000000in}{-0.034722in}}%
\pgfpathlineto{\pgfqpoint{0.000000in}{0.034722in}}%
\pgfusepath{stroke,fill}%
}%
\begin{pgfscope}%
\pgfsys@transformshift{1.224444in}{2.112000in}%
\pgfsys@useobject{currentmarker}{}%
\end{pgfscope}%
\end{pgfscope}%
\begin{pgfscope}%
\pgftext[x=1.524444in,y=2.053667in,left,base]{\sffamily\fontsize{14.000000}{14.400000}\selectfont B-UP = 5}%
\end{pgfscope}%
\end{pgfpicture}%
\makeatother%
\endgroup%

%% file: figures/dpm_variants.pgf
\begingroup%
\makeatletter%
\begin{pgfpicture}%
\pgfpathrectangle{\pgfpointorigin}{\pgfqpoint{5.000000in}{3.000000in}}%
\pgfusepath{use as bounding box, clip}%
\begin{pgfscope}%
\pgfsetbuttcap%
\pgfsetmiterjoin%
\definecolor{currentfill}{rgb}{1.000000,1.000000,1.000000}%
\pgfsetfillcolor{currentfill}%
\pgfsetlinewidth{0.000000pt}%
\definecolor{currentstroke}{rgb}{1.000000,1.000000,1.000000}%
\pgfsetstrokecolor{currentstroke}%
\pgfsetdash{}{0pt}%
\pgfpathmoveto{\pgfqpoint{0.000000in}{0.000000in}}%
\pgfpathlineto{\pgfqpoint{5.000000in}{0.000000in}}%
\pgfpathlineto{\pgfqpoint{5.000000in}{3.000000in}}%
\pgfpathlineto{\pgfqpoint{0.000000in}{3.000000in}}%
\pgfpathclose%
\pgfusepath{fill}%
\end{pgfscope}%
\begin{pgfscope}%
\pgfsetbuttcap%
\pgfsetmiterjoin%
\definecolor{currentfill}{rgb}{1.000000,1.000000,1.000000}%
\pgfsetfillcolor{currentfill}%
\pgfsetlinewidth{0.000000pt}%
\definecolor{currentstroke}{rgb}{0.000000,0.000000,0.000000}%
\pgfsetstrokecolor{currentstroke}%
\pgfsetstrokeopacity{0.000000}%
\pgfsetdash{}{0pt}%
\pgfpathmoveto{\pgfqpoint{0.801528in}{0.632778in}}%
\pgfpathlineto{\pgfqpoint{4.092222in}{0.632778in}}%
\pgfpathlineto{\pgfqpoint{4.092222in}{2.785000in}}%
\pgfpathlineto{\pgfqpoint{0.801528in}{2.785000in}}%
\pgfpathclose%
\pgfusepath{fill}%
\end{pgfscope}%
\begin{pgfscope}%
\pgfsetbuttcap%
\pgfsetroundjoin%
\definecolor{currentfill}{rgb}{0.000000,0.000000,0.000000}%
\pgfsetfillcolor{currentfill}%
\pgfsetlinewidth{0.803000pt}%
\definecolor{currentstroke}{rgb}{0.000000,0.000000,0.000000}%
\pgfsetstrokecolor{currentstroke}%
\pgfsetdash{}{0pt}%
\pgfsys@defobject{currentmarker}{\pgfqpoint{0.000000in}{-0.048611in}}{\pgfqpoint{0.000000in}{0.000000in}}{%
\pgfpathmoveto{\pgfqpoint{0.000000in}{0.000000in}}%
\pgfpathlineto{\pgfqpoint{0.000000in}{-0.048611in}}%
\pgfusepath{stroke,fill}%
}%
\begin{pgfscope}%
\pgfsys@transformshift{0.951105in}{0.632778in}%
\pgfsys@useobject{currentmarker}{}%
\end{pgfscope}%
\end{pgfscope}%
\begin{pgfscope}%
\pgftext[x=0.951105in,y=0.535556in,,top]{\sffamily\fontsize{16.000000}{14.400000}\selectfont Sk1}%
\end{pgfscope}%
\begin{pgfscope}%
\pgfsetbuttcap%
\pgfsetroundjoin%
\definecolor{currentfill}{rgb}{0.000000,0.000000,0.000000}%
\pgfsetfillcolor{currentfill}%
\pgfsetlinewidth{0.803000pt}%
\definecolor{currentstroke}{rgb}{0.000000,0.000000,0.000000}%
\pgfsetstrokecolor{currentstroke}%
\pgfsetdash{}{0pt}%
\pgfsys@defobject{currentmarker}{\pgfqpoint{0.000000in}{-0.048611in}}{\pgfqpoint{0.000000in}{0.000000in}}{%
\pgfpathmoveto{\pgfqpoint{0.000000in}{0.000000in}}%
\pgfpathlineto{\pgfqpoint{0.000000in}{-0.048611in}}%
\pgfusepath{stroke,fill}%
}%
\begin{pgfscope}%
\pgfsys@transformshift{1.948285in}{0.632778in}%
\pgfsys@useobject{currentmarker}{}%
\end{pgfscope}%
\end{pgfscope}%
\begin{pgfscope}%
\pgftext[x=1.948285in,y=0.535556in,,top]{\sffamily\fontsize{16.000000}{14.400000}\selectfont Sk2}%
\end{pgfscope}%
\begin{pgfscope}%
\pgfsetbuttcap%
\pgfsetroundjoin%
\definecolor{currentfill}{rgb}{0.000000,0.000000,0.000000}%
\pgfsetfillcolor{currentfill}%
\pgfsetlinewidth{0.803000pt}%
\definecolor{currentstroke}{rgb}{0.000000,0.000000,0.000000}%
\pgfsetstrokecolor{currentstroke}%
\pgfsetdash{}{0pt}%
\pgfsys@defobject{currentmarker}{\pgfqpoint{0.000000in}{-0.048611in}}{\pgfqpoint{0.000000in}{0.000000in}}{%
\pgfpathmoveto{\pgfqpoint{0.000000in}{0.000000in}}%
\pgfpathlineto{\pgfqpoint{0.000000in}{-0.048611in}}%
\pgfusepath{stroke,fill}%
}%
\begin{pgfscope}%
\pgfsys@transformshift{2.945465in}{0.632778in}%
\pgfsys@useobject{currentmarker}{}%
\end{pgfscope}%
\end{pgfscope}%
\begin{pgfscope}%
\pgftext[x=2.945465in,y=0.535556in,,top]{\sffamily\fontsize{16.000000}{14.400000}\selectfont Sk3}%
\end{pgfscope}%
\begin{pgfscope}%
\pgfsetbuttcap%
\pgfsetroundjoin%
\definecolor{currentfill}{rgb}{0.000000,0.000000,0.000000}%
\pgfsetfillcolor{currentfill}%
\pgfsetlinewidth{0.803000pt}%
\definecolor{currentstroke}{rgb}{0.000000,0.000000,0.000000}%
\pgfsetstrokecolor{currentstroke}%
\pgfsetdash{}{0pt}%
\pgfsys@defobject{currentmarker}{\pgfqpoint{0.000000in}{-0.048611in}}{\pgfqpoint{0.000000in}{0.000000in}}{%
\pgfpathmoveto{\pgfqpoint{0.000000in}{0.000000in}}%
\pgfpathlineto{\pgfqpoint{0.000000in}{-0.048611in}}%
\pgfusepath{stroke,fill}%
}%
\begin{pgfscope}%
\pgfsys@transformshift{3.942645in}{0.632778in}%
\pgfsys@useobject{currentmarker}{}%
\end{pgfscope}%
\end{pgfscope}%
\begin{pgfscope}%
\pgftext[x=3.942645in,y=0.535556in,,top]{\sffamily\fontsize{16.000000}{14.400000}\selectfont Sk4}%
\end{pgfscope}%
\begin{pgfscope}%
\pgftext[x=2.446875in,y=0.332000in,,top]{\sffamily\fontsize{16.000000}{14.400000}\selectfont variants of sketch}%
\end{pgfscope}%
\begin{pgfscope}%
\pgfsetbuttcap%
\pgfsetroundjoin%
\definecolor{currentfill}{rgb}{0.000000,0.000000,0.000000}%
\pgfsetfillcolor{currentfill}%
\pgfsetlinewidth{0.803000pt}%
\definecolor{currentstroke}{rgb}{0.000000,0.000000,0.000000}%
\pgfsetstrokecolor{currentstroke}%
\pgfsetdash{}{0pt}%
\pgfsys@defobject{currentmarker}{\pgfqpoint{-0.048611in}{0.000000in}}{\pgfqpoint{0.000000in}{0.000000in}}{%
\pgfpathmoveto{\pgfqpoint{0.000000in}{0.000000in}}%
\pgfpathlineto{\pgfqpoint{-0.048611in}{0.000000in}}%
\pgfusepath{stroke,fill}%
}%
\begin{pgfscope}%
\pgfsys@transformshift{0.801528in}{0.632778in}%
\pgfsys@useobject{currentmarker}{}%
\end{pgfscope}%
\end{pgfscope}%
\begin{pgfscope}%
\pgftext[x=0.622639in,y=0.574944in,left,base]{\sffamily\fontsize{16.000000}{14.400000}\selectfont 0}%
\end{pgfscope}%
\begin{pgfscope}%
\pgfsetbuttcap%
\pgfsetroundjoin%
\definecolor{currentfill}{rgb}{0.000000,0.000000,0.000000}%
\pgfsetfillcolor{currentfill}%
\pgfsetlinewidth{0.803000pt}%
\definecolor{currentstroke}{rgb}{0.000000,0.000000,0.000000}%
\pgfsetstrokecolor{currentstroke}%
\pgfsetdash{}{0pt}%
\pgfsys@defobject{currentmarker}{\pgfqpoint{-0.048611in}{0.000000in}}{\pgfqpoint{0.000000in}{0.000000in}}{%
\pgfpathmoveto{\pgfqpoint{0.000000in}{0.000000in}}%
\pgfpathlineto{\pgfqpoint{-0.048611in}{0.000000in}}%
\pgfusepath{stroke,fill}%
}%
\begin{pgfscope}%
\pgfsys@transformshift{0.801528in}{1.130862in}%
\pgfsys@useobject{currentmarker}{}%
\end{pgfscope}%
\end{pgfscope}%
\begin{pgfscope}%
\pgftext[x=0.470972in,y=1.073029in,left,base]{\sffamily\fontsize{16.000000}{14.400000}\selectfont 50}%
\end{pgfscope}%
\begin{pgfscope}%
\pgfsetbuttcap%
\pgfsetroundjoin%
\definecolor{currentfill}{rgb}{0.000000,0.000000,0.000000}%
\pgfsetfillcolor{currentfill}%
\pgfsetlinewidth{0.803000pt}%
\definecolor{currentstroke}{rgb}{0.000000,0.000000,0.000000}%
\pgfsetstrokecolor{currentstroke}%
\pgfsetdash{}{0pt}%
\pgfsys@defobject{currentmarker}{\pgfqpoint{-0.048611in}{0.000000in}}{\pgfqpoint{0.000000in}{0.000000in}}{%
\pgfpathmoveto{\pgfqpoint{0.000000in}{0.000000in}}%
\pgfpathlineto{\pgfqpoint{-0.048611in}{0.000000in}}%
\pgfusepath{stroke,fill}%
}%
\begin{pgfscope}%
\pgfsys@transformshift{0.801528in}{1.628946in}%
\pgfsys@useobject{currentmarker}{}%
\end{pgfscope}%
\end{pgfscope}%
\begin{pgfscope}%
\pgftext[x=0.389306in,y=1.571113in,left,base]{\sffamily\fontsize{16.000000}{14.400000}\selectfont 100}%
\end{pgfscope}%
\begin{pgfscope}%
\pgfsetbuttcap%
\pgfsetroundjoin%
\definecolor{currentfill}{rgb}{0.000000,0.000000,0.000000}%
\pgfsetfillcolor{currentfill}%
\pgfsetlinewidth{0.803000pt}%
\definecolor{currentstroke}{rgb}{0.000000,0.000000,0.000000}%
\pgfsetstrokecolor{currentstroke}%
\pgfsetdash{}{0pt}%
\pgfsys@defobject{currentmarker}{\pgfqpoint{-0.048611in}{0.000000in}}{\pgfqpoint{0.000000in}{0.000000in}}{%
\pgfpathmoveto{\pgfqpoint{0.000000in}{0.000000in}}%
\pgfpathlineto{\pgfqpoint{-0.048611in}{0.000000in}}%
\pgfusepath{stroke,fill}%
}%
\begin{pgfscope}%
\pgfsys@transformshift{0.801528in}{2.127031in}%
\pgfsys@useobject{currentmarker}{}%
\end{pgfscope}%
\end{pgfscope}%
\begin{pgfscope}%
\pgftext[x=0.389306in,y=2.069197in,left,base]{\sffamily\fontsize{16.000000}{14.400000}\selectfont 150}%
\end{pgfscope}%
\begin{pgfscope}%
\pgfsetbuttcap%
\pgfsetroundjoin%
\definecolor{currentfill}{rgb}{0.000000,0.000000,0.000000}%
\pgfsetfillcolor{currentfill}%
\pgfsetlinewidth{0.803000pt}%
\definecolor{currentstroke}{rgb}{0.000000,0.000000,0.000000}%
\pgfsetstrokecolor{currentstroke}%
\pgfsetdash{}{0pt}%
\pgfsys@defobject{currentmarker}{\pgfqpoint{-0.048611in}{0.000000in}}{\pgfqpoint{0.000000in}{0.000000in}}{%
\pgfpathmoveto{\pgfqpoint{0.000000in}{0.000000in}}%
\pgfpathlineto{\pgfqpoint{-0.048611in}{0.000000in}}%
\pgfusepath{stroke,fill}%
}%
\begin{pgfscope}%
\pgfsys@transformshift{0.801528in}{2.625115in}%
\pgfsys@useobject{currentmarker}{}%
\end{pgfscope}%
\end{pgfscope}%
\begin{pgfscope}%
\pgftext[x=0.389306in,y=2.567282in,left,base]{\sffamily\fontsize{16.000000}{14.400000}\selectfont 200}%
\end{pgfscope}%
\begin{pgfscope}%
\definecolor{textcolor}{rgb}{0.000000,0.500000,0.000000}%
\pgfsetstrokecolor{textcolor}%
\pgfsetfillcolor{textcolor}%
\pgftext[x=0.323750in,y=1.708889in,,bottom,rotate=90.000000]{\color{textcolor}\sffamily\fontsize{16.000000}{14.400000}\selectfont iterations}%
\end{pgfscope}%
\begin{pgfscope}%
\pgfpathrectangle{\pgfqpoint{0.801528in}{0.632778in}}{\pgfqpoint{3.290694in}{2.152222in}} %
\pgfusepath{clip}%
\pgfsetbuttcap%
\pgfsetroundjoin%
\definecolor{currentfill}{rgb}{0.000000,0.500000,0.000000}%
\pgfsetfillcolor{currentfill}%
\pgfsetlinewidth{1.003750pt}%
\definecolor{currentstroke}{rgb}{0.000000,0.500000,0.000000}%
\pgfsetstrokecolor{currentstroke}%
\pgfsetdash{}{0pt}%
\pgfsys@defobject{currentmarker}{\pgfqpoint{-0.041667in}{-0.041667in}}{\pgfqpoint{0.041667in}{0.041667in}}{%
\pgfpathmoveto{\pgfqpoint{0.000000in}{-0.041667in}}%
\pgfpathcurveto{\pgfqpoint{0.011050in}{-0.041667in}}{\pgfqpoint{0.021649in}{-0.037276in}}{\pgfqpoint{0.029463in}{-0.029463in}}%
\pgfpathcurveto{\pgfqpoint{0.037276in}{-0.021649in}}{\pgfqpoint{0.041667in}{-0.011050in}}{\pgfqpoint{0.041667in}{0.000000in}}%
\pgfpathcurveto{\pgfqpoint{0.041667in}{0.011050in}}{\pgfqpoint{0.037276in}{0.021649in}}{\pgfqpoint{0.029463in}{0.029463in}}%
\pgfpathcurveto{\pgfqpoint{0.021649in}{0.037276in}}{\pgfqpoint{0.011050in}{0.041667in}}{\pgfqpoint{0.000000in}{0.041667in}}%
\pgfpathcurveto{\pgfqpoint{-0.011050in}{0.041667in}}{\pgfqpoint{-0.021649in}{0.037276in}}{\pgfqpoint{-0.029463in}{0.029463in}}%
\pgfpathcurveto{\pgfqpoint{-0.037276in}{0.021649in}}{\pgfqpoint{-0.041667in}{0.011050in}}{\pgfqpoint{-0.041667in}{0.000000in}}%
\pgfpathcurveto{\pgfqpoint{-0.041667in}{-0.011050in}}{\pgfqpoint{-0.037276in}{-0.021649in}}{\pgfqpoint{-0.029463in}{-0.029463in}}%
\pgfpathcurveto{\pgfqpoint{-0.021649in}{-0.037276in}}{\pgfqpoint{-0.011050in}{-0.041667in}}{\pgfqpoint{0.000000in}{-0.041667in}}%
\pgfpathclose%
\pgfusepath{stroke,fill}%
}%
\begin{pgfscope}%
\pgfsys@transformshift{0.951105in}{2.127031in}%
\pgfsys@useobject{currentmarker}{}%
\end{pgfscope}%
\begin{pgfscope}%
\pgfsys@transformshift{1.948285in}{0.891782in}%
\pgfsys@useobject{currentmarker}{}%
\end{pgfscope}%
\begin{pgfscope}%
\pgfsys@transformshift{2.945465in}{2.694847in}%
\pgfsys@useobject{currentmarker}{}%
\end{pgfscope}%
\begin{pgfscope}%
\pgfsys@transformshift{3.942645in}{1.907874in}%
\pgfsys@useobject{currentmarker}{}%
\end{pgfscope}%
\end{pgfscope}%
\begin{pgfscope}%
\pgfpathrectangle{\pgfqpoint{0.801528in}{0.632778in}}{\pgfqpoint{3.290694in}{2.152222in}} %
\pgfusepath{clip}%
\pgfsetbuttcap%
\pgfsetroundjoin%
\definecolor{currentfill}{rgb}{0.000000,0.500000,0.000000}%
\pgfsetfillcolor{currentfill}%
\pgfsetlinewidth{1.003750pt}%
\definecolor{currentstroke}{rgb}{0.000000,0.500000,0.000000}%
\pgfsetstrokecolor{currentstroke}%
\pgfsetdash{}{0pt}%
\pgfsys@defobject{currentmarker}{\pgfqpoint{-0.041667in}{-0.041667in}}{\pgfqpoint{0.041667in}{0.041667in}}{%
\pgfpathmoveto{\pgfqpoint{-0.041667in}{-0.041667in}}%
\pgfpathlineto{\pgfqpoint{0.041667in}{0.041667in}}%
\pgfpathmoveto{\pgfqpoint{-0.041667in}{0.041667in}}%
\pgfpathlineto{\pgfqpoint{0.041667in}{-0.041667in}}%
\pgfusepath{stroke,fill}%
}%
\begin{pgfscope}%
\pgfsys@transformshift{0.951105in}{2.127031in}%
\pgfsys@useobject{currentmarker}{}%
\end{pgfscope}%
\begin{pgfscope}%
\pgfsys@transformshift{1.948285in}{0.891782in}%
\pgfsys@useobject{currentmarker}{}%
\end{pgfscope}%
\begin{pgfscope}%
\pgfsys@transformshift{2.945465in}{2.674923in}%
\pgfsys@useobject{currentmarker}{}%
\end{pgfscope}%
\begin{pgfscope}%
\pgfsys@transformshift{3.942645in}{1.907874in}%
\pgfsys@useobject{currentmarker}{}%
\end{pgfscope}%
\end{pgfscope}%
\begin{pgfscope}%
\pgfsetrectcap%
\pgfsetmiterjoin%
\pgfsetlinewidth{0.803000pt}%
\definecolor{currentstroke}{rgb}{0.000000,0.000000,0.000000}%
\pgfsetstrokecolor{currentstroke}%
\pgfsetdash{}{0pt}%
\pgfpathmoveto{\pgfqpoint{0.801528in}{0.632778in}}%
\pgfpathlineto{\pgfqpoint{0.801528in}{2.785000in}}%
\pgfusepath{stroke}%
\end{pgfscope}%
\begin{pgfscope}%
\pgfsetrectcap%
\pgfsetmiterjoin%
\pgfsetlinewidth{0.803000pt}%
\definecolor{currentstroke}{rgb}{0.000000,0.000000,0.000000}%
\pgfsetstrokecolor{currentstroke}%
\pgfsetdash{}{0pt}%
\pgfpathmoveto{\pgfqpoint{4.092222in}{0.632778in}}%
\pgfpathlineto{\pgfqpoint{4.092222in}{2.785000in}}%
\pgfusepath{stroke}%
\end{pgfscope}%
\begin{pgfscope}%
\pgfsetrectcap%
\pgfsetmiterjoin%
\pgfsetlinewidth{0.803000pt}%
\definecolor{currentstroke}{rgb}{0.000000,0.000000,0.000000}%
\pgfsetstrokecolor{currentstroke}%
\pgfsetdash{}{0pt}%
\pgfpathmoveto{\pgfqpoint{0.801528in}{0.632778in}}%
\pgfpathlineto{\pgfqpoint{4.092222in}{0.632778in}}%
\pgfusepath{stroke}%
\end{pgfscope}%
\begin{pgfscope}%
\pgfsetrectcap%
\pgfsetmiterjoin%
\pgfsetlinewidth{0.803000pt}%
\definecolor{currentstroke}{rgb}{0.000000,0.000000,0.000000}%
\pgfsetstrokecolor{currentstroke}%
\pgfsetdash{}{0pt}%
\pgfpathmoveto{\pgfqpoint{0.801528in}{2.785000in}}%
\pgfpathlineto{\pgfqpoint{4.092222in}{2.785000in}}%
\pgfusepath{stroke}%
\end{pgfscope}%
\begin{pgfscope}%
\pgfsetbuttcap%
\pgfsetroundjoin%
\definecolor{currentfill}{rgb}{0.000000,0.000000,0.000000}%
\pgfsetfillcolor{currentfill}%
\pgfsetlinewidth{0.803000pt}%
\definecolor{currentstroke}{rgb}{0.000000,0.000000,0.000000}%
\pgfsetstrokecolor{currentstroke}%
\pgfsetdash{}{0pt}%
\pgfsys@defobject{currentmarker}{\pgfqpoint{0.000000in}{0.000000in}}{\pgfqpoint{0.048611in}{0.000000in}}{%
\pgfpathmoveto{\pgfqpoint{0.000000in}{0.000000in}}%
\pgfpathlineto{\pgfqpoint{0.048611in}{0.000000in}}%
\pgfusepath{stroke,fill}%
}%
\begin{pgfscope}%
\pgfsys@transformshift{4.092222in}{0.632778in}%
\pgfsys@useobject{currentmarker}{}%
\end{pgfscope}%
\end{pgfscope}%
\begin{pgfscope}%
\pgftext[x=4.189444in,y=0.574944in,left,base]{\sffamily\fontsize{16.000000}{14.400000}\selectfont 0}%
\end{pgfscope}%
\begin{pgfscope}%
\pgfsetbuttcap%
\pgfsetroundjoin%
\definecolor{currentfill}{rgb}{0.000000,0.000000,0.000000}%
\pgfsetfillcolor{currentfill}%
\pgfsetlinewidth{0.803000pt}%
\definecolor{currentstroke}{rgb}{0.000000,0.000000,0.000000}%
\pgfsetstrokecolor{currentstroke}%
\pgfsetdash{}{0pt}%
\pgfsys@defobject{currentmarker}{\pgfqpoint{0.000000in}{0.000000in}}{\pgfqpoint{0.048611in}{0.000000in}}{%
\pgfpathmoveto{\pgfqpoint{0.000000in}{0.000000in}}%
\pgfpathlineto{\pgfqpoint{0.048611in}{0.000000in}}%
\pgfusepath{stroke,fill}%
}%
\begin{pgfscope}%
\pgfsys@transformshift{4.092222in}{1.009432in}%
\pgfsys@useobject{currentmarker}{}%
\end{pgfscope}%
\end{pgfscope}%
\begin{pgfscope}%
\pgftext[x=4.189444in,y=0.951599in,left,base]{\sffamily\fontsize{16.000000}{14.400000}\selectfont 200}%
\end{pgfscope}%
\begin{pgfscope}%
\pgfsetbuttcap%
\pgfsetroundjoin%
\definecolor{currentfill}{rgb}{0.000000,0.000000,0.000000}%
\pgfsetfillcolor{currentfill}%
\pgfsetlinewidth{0.803000pt}%
\definecolor{currentstroke}{rgb}{0.000000,0.000000,0.000000}%
\pgfsetstrokecolor{currentstroke}%
\pgfsetdash{}{0pt}%
\pgfsys@defobject{currentmarker}{\pgfqpoint{0.000000in}{0.000000in}}{\pgfqpoint{0.048611in}{0.000000in}}{%
\pgfpathmoveto{\pgfqpoint{0.000000in}{0.000000in}}%
\pgfpathlineto{\pgfqpoint{0.048611in}{0.000000in}}%
\pgfusepath{stroke,fill}%
}%
\begin{pgfscope}%
\pgfsys@transformshift{4.092222in}{1.386087in}%
\pgfsys@useobject{currentmarker}{}%
\end{pgfscope}%
\end{pgfscope}%
\begin{pgfscope}%
\pgftext[x=4.189444in,y=1.328253in,left,base]{\sffamily\fontsize{16.000000}{14.400000}\selectfont 400}%
\end{pgfscope}%
\begin{pgfscope}%
\pgfsetbuttcap%
\pgfsetroundjoin%
\definecolor{currentfill}{rgb}{0.000000,0.000000,0.000000}%
\pgfsetfillcolor{currentfill}%
\pgfsetlinewidth{0.803000pt}%
\definecolor{currentstroke}{rgb}{0.000000,0.000000,0.000000}%
\pgfsetstrokecolor{currentstroke}%
\pgfsetdash{}{0pt}%
\pgfsys@defobject{currentmarker}{\pgfqpoint{0.000000in}{0.000000in}}{\pgfqpoint{0.048611in}{0.000000in}}{%
\pgfpathmoveto{\pgfqpoint{0.000000in}{0.000000in}}%
\pgfpathlineto{\pgfqpoint{0.048611in}{0.000000in}}%
\pgfusepath{stroke,fill}%
}%
\begin{pgfscope}%
\pgfsys@transformshift{4.092222in}{1.762741in}%
\pgfsys@useobject{currentmarker}{}%
\end{pgfscope}%
\end{pgfscope}%
\begin{pgfscope}%
\pgftext[x=4.189444in,y=1.704908in,left,base]{\sffamily\fontsize{16.000000}{14.400000}\selectfont 600}%
\end{pgfscope}%
\begin{pgfscope}%
\pgfsetbuttcap%
\pgfsetroundjoin%
\definecolor{currentfill}{rgb}{0.000000,0.000000,0.000000}%
\pgfsetfillcolor{currentfill}%
\pgfsetlinewidth{0.803000pt}%
\definecolor{currentstroke}{rgb}{0.000000,0.000000,0.000000}%
\pgfsetstrokecolor{currentstroke}%
\pgfsetdash{}{0pt}%
\pgfsys@defobject{currentmarker}{\pgfqpoint{0.000000in}{0.000000in}}{\pgfqpoint{0.048611in}{0.000000in}}{%
\pgfpathmoveto{\pgfqpoint{0.000000in}{0.000000in}}%
\pgfpathlineto{\pgfqpoint{0.048611in}{0.000000in}}%
\pgfusepath{stroke,fill}%
}%
\begin{pgfscope}%
\pgfsys@transformshift{4.092222in}{2.139395in}%
\pgfsys@useobject{currentmarker}{}%
\end{pgfscope}%
\end{pgfscope}%
\begin{pgfscope}%
\pgftext[x=4.189444in,y=2.081562in,left,base]{\sffamily\fontsize{16.000000}{14.400000}\selectfont 800}%
\end{pgfscope}%
\begin{pgfscope}%
\pgfsetbuttcap%
\pgfsetroundjoin%
\definecolor{currentfill}{rgb}{0.000000,0.000000,0.000000}%
\pgfsetfillcolor{currentfill}%
\pgfsetlinewidth{0.803000pt}%
\definecolor{currentstroke}{rgb}{0.000000,0.000000,0.000000}%
\pgfsetstrokecolor{currentstroke}%
\pgfsetdash{}{0pt}%
\pgfsys@defobject{currentmarker}{\pgfqpoint{0.000000in}{0.000000in}}{\pgfqpoint{0.048611in}{0.000000in}}{%
\pgfpathmoveto{\pgfqpoint{0.000000in}{0.000000in}}%
\pgfpathlineto{\pgfqpoint{0.048611in}{0.000000in}}%
\pgfusepath{stroke,fill}%
}%
\begin{pgfscope}%
\pgfsys@transformshift{4.092222in}{2.516050in}%
\pgfsys@useobject{currentmarker}{}%
\end{pgfscope}%
\end{pgfscope}%
\begin{pgfscope}%
\pgftext[x=4.189444in,y=2.458216in,left,base]{\sffamily\fontsize{16.000000}{14.400000}\selectfont 1000}%
\end{pgfscope}%
\begin{pgfscope}%
\definecolor{textcolor}{rgb}{0.000000,0.000000,1.000000}%
\pgfsetstrokecolor{textcolor}%
\pgfsetfillcolor{textcolor}%
\pgftext[x=4.671667in,y=1.708889in,,top,rotate=90.000000]{\color{textcolor}\sffamily\fontsize{16.000000}{14.400000}\selectfont time (seconds)}%
\end{pgfscope}%
\begin{pgfscope}%
\pgfpathrectangle{\pgfqpoint{0.801528in}{0.632778in}}{\pgfqpoint{3.290694in}{2.152222in}} %
\pgfusepath{clip}%
\pgfsetbuttcap%
\pgfsetroundjoin%
\definecolor{currentfill}{rgb}{0.000000,0.000000,1.000000}%
\pgfsetfillcolor{currentfill}%
\pgfsetlinewidth{1.003750pt}%
\definecolor{currentstroke}{rgb}{0.000000,0.000000,1.000000}%
\pgfsetstrokecolor{currentstroke}%
\pgfsetdash{}{0pt}%
\pgfsys@defobject{currentmarker}{\pgfqpoint{-0.034722in}{-0.034722in}}{\pgfqpoint{0.034722in}{0.034722in}}{%
\pgfpathmoveto{\pgfqpoint{0.000000in}{-0.034722in}}%
\pgfpathcurveto{\pgfqpoint{0.009208in}{-0.034722in}}{\pgfqpoint{0.018041in}{-0.031064in}}{\pgfqpoint{0.024552in}{-0.024552in}}%
\pgfpathcurveto{\pgfqpoint{0.031064in}{-0.018041in}}{\pgfqpoint{0.034722in}{-0.009208in}}{\pgfqpoint{0.034722in}{0.000000in}}%
\pgfpathcurveto{\pgfqpoint{0.034722in}{0.009208in}}{\pgfqpoint{0.031064in}{0.018041in}}{\pgfqpoint{0.024552in}{0.024552in}}%
\pgfpathcurveto{\pgfqpoint{0.018041in}{0.031064in}}{\pgfqpoint{0.009208in}{0.034722in}}{\pgfqpoint{0.000000in}{0.034722in}}%
\pgfpathcurveto{\pgfqpoint{-0.009208in}{0.034722in}}{\pgfqpoint{-0.018041in}{0.031064in}}{\pgfqpoint{-0.024552in}{0.024552in}}%
\pgfpathcurveto{\pgfqpoint{-0.031064in}{0.018041in}}{\pgfqpoint{-0.034722in}{0.009208in}}{\pgfqpoint{-0.034722in}{0.000000in}}%
\pgfpathcurveto{\pgfqpoint{-0.034722in}{-0.009208in}}{\pgfqpoint{-0.031064in}{-0.018041in}}{\pgfqpoint{-0.024552in}{-0.024552in}}%
\pgfpathcurveto{\pgfqpoint{-0.018041in}{-0.031064in}}{\pgfqpoint{-0.009208in}{-0.034722in}}{\pgfqpoint{0.000000in}{-0.034722in}}%
\pgfpathclose%
\pgfusepath{stroke,fill}%
}%
\begin{pgfscope}%
\pgfsys@transformshift{0.951105in}{0.972931in}%
\pgfsys@useobject{currentmarker}{}%
\end{pgfscope}%
\begin{pgfscope}%
\pgfsys@transformshift{1.948285in}{0.675476in}%
\pgfsys@useobject{currentmarker}{}%
\end{pgfscope}%
\begin{pgfscope}%
\pgfsys@transformshift{2.945465in}{1.011293in}%
\pgfsys@useobject{currentmarker}{}%
\end{pgfscope}%
\begin{pgfscope}%
\pgfsys@transformshift{3.942645in}{0.861984in}%
\pgfsys@useobject{currentmarker}{}%
\end{pgfscope}%
\end{pgfscope}%
\begin{pgfscope}%
\pgfpathrectangle{\pgfqpoint{0.801528in}{0.632778in}}{\pgfqpoint{3.290694in}{2.152222in}} %
\pgfusepath{clip}%
\pgfsetbuttcap%
\pgfsetroundjoin%
\definecolor{currentfill}{rgb}{0.000000,0.000000,1.000000}%
\pgfsetfillcolor{currentfill}%
\pgfsetlinewidth{1.003750pt}%
\definecolor{currentstroke}{rgb}{0.000000,0.000000,1.000000}%
\pgfsetstrokecolor{currentstroke}%
\pgfsetdash{}{0pt}%
\pgfsys@defobject{currentmarker}{\pgfqpoint{-0.034722in}{-0.034722in}}{\pgfqpoint{0.034722in}{0.034722in}}{%
\pgfpathmoveto{\pgfqpoint{-0.034722in}{-0.034722in}}%
\pgfpathlineto{\pgfqpoint{0.034722in}{0.034722in}}%
\pgfpathmoveto{\pgfqpoint{-0.034722in}{0.034722in}}%
\pgfpathlineto{\pgfqpoint{0.034722in}{-0.034722in}}%
\pgfusepath{stroke,fill}%
}%
\begin{pgfscope}%
\pgfsys@transformshift{0.951105in}{2.684546in}%
\pgfsys@useobject{currentmarker}{}%
\end{pgfscope}%
\begin{pgfscope}%
\pgfsys@transformshift{1.948285in}{0.816143in}%
\pgfsys@useobject{currentmarker}{}%
\end{pgfscope}%
\begin{pgfscope}%
\pgfsys@transformshift{2.945465in}{2.675295in}%
\pgfsys@useobject{currentmarker}{}%
\end{pgfscope}%
\begin{pgfscope}%
\pgfsys@transformshift{3.942645in}{1.686974in}%
\pgfsys@useobject{currentmarker}{}%
\end{pgfscope}%
\end{pgfscope}%
\begin{pgfscope}%
\pgfsetrectcap%
\pgfsetmiterjoin%
\pgfsetlinewidth{0.803000pt}%
\definecolor{currentstroke}{rgb}{0.000000,0.000000,0.000000}%
\pgfsetstrokecolor{currentstroke}%
\pgfsetdash{}{0pt}%
\pgfpathmoveto{\pgfqpoint{0.801528in}{0.632778in}}%
\pgfpathlineto{\pgfqpoint{0.801528in}{2.785000in}}%
\pgfusepath{stroke}%
\end{pgfscope}%
\begin{pgfscope}%
\pgfsetrectcap%
\pgfsetmiterjoin%
\pgfsetlinewidth{0.803000pt}%
\definecolor{currentstroke}{rgb}{0.000000,0.000000,0.000000}%
\pgfsetstrokecolor{currentstroke}%
\pgfsetdash{}{0pt}%
\pgfpathmoveto{\pgfqpoint{4.092222in}{0.632778in}}%
\pgfpathlineto{\pgfqpoint{4.092222in}{2.785000in}}%
\pgfusepath{stroke}%
\end{pgfscope}%
\begin{pgfscope}%
\pgfsetrectcap%
\pgfsetmiterjoin%
\pgfsetlinewidth{0.803000pt}%
\definecolor{currentstroke}{rgb}{0.000000,0.000000,0.000000}%
\pgfsetstrokecolor{currentstroke}%
\pgfsetdash{}{0pt}%
\pgfpathmoveto{\pgfqpoint{0.801528in}{0.632778in}}%
\pgfpathlineto{\pgfqpoint{4.092222in}{0.632778in}}%
\pgfusepath{stroke}%
\end{pgfscope}%
\begin{pgfscope}%
\pgfsetrectcap%
\pgfsetmiterjoin%
\pgfsetlinewidth{0.803000pt}%
\definecolor{currentstroke}{rgb}{0.000000,0.000000,0.000000}%
\pgfsetstrokecolor{currentstroke}%
\pgfsetdash{}{0pt}%
\pgfpathmoveto{\pgfqpoint{0.801528in}{2.785000in}}%
\pgfpathlineto{\pgfqpoint{4.092222in}{2.785000in}}%
\pgfusepath{stroke}%
\end{pgfscope}%
\begin{pgfscope}%
\pgfsetbuttcap%
\pgfsetmiterjoin%
\definecolor{currentfill}{rgb}{0.300000,0.300000,0.300000}%
\pgfsetfillcolor{currentfill}%
\pgfsetfillopacity{0.500000}%
\pgfsetlinewidth{1.003750pt}%
\definecolor{currentstroke}{rgb}{0.300000,0.300000,0.300000}%
\pgfsetstrokecolor{currentstroke}%
\pgfsetstrokeopacity{0.500000}%
\pgfsetdash{}{0pt}%
\pgfpathmoveto{\pgfqpoint{0.945972in}{1.423778in}}%
\pgfpathlineto{\pgfqpoint{2.358306in}{1.423778in}}%
\pgfpathquadraticcurveto{\pgfqpoint{2.391639in}{1.423778in}}{\pgfqpoint{2.391639in}{1.457111in}}%
\pgfpathlineto{\pgfqpoint{2.391639in}{1.905111in}}%
\pgfpathquadraticcurveto{\pgfqpoint{2.391639in}{1.938444in}}{\pgfqpoint{2.358306in}{1.938444in}}%
\pgfpathlineto{\pgfqpoint{0.945972in}{1.938444in}}%
\pgfpathquadraticcurveto{\pgfqpoint{0.912639in}{1.938444in}}{\pgfqpoint{0.912639in}{1.905111in}}%
\pgfpathlineto{\pgfqpoint{0.912639in}{1.457111in}}%
\pgfpathquadraticcurveto{\pgfqpoint{0.912639in}{1.423778in}}{\pgfqpoint{0.945972in}{1.423778in}}%
\pgfpathclose%
\pgfusepath{stroke,fill}%
\end{pgfscope}%
\begin{pgfscope}%
\pgfsetbuttcap%
\pgfsetmiterjoin%
\definecolor{currentfill}{rgb}{1.000000,1.000000,1.000000}%
\pgfsetfillcolor{currentfill}%
\pgfsetfillopacity{0.500000}%
\pgfsetlinewidth{1.003750pt}%
\definecolor{currentstroke}{rgb}{0.800000,0.800000,0.800000}%
\pgfsetstrokecolor{currentstroke}%
\pgfsetstrokeopacity{0.500000}%
\pgfsetdash{}{0pt}%
\pgfpathmoveto{\pgfqpoint{0.918194in}{1.451556in}}%
\pgfpathlineto{\pgfqpoint{2.330528in}{1.451556in}}%
\pgfpathquadraticcurveto{\pgfqpoint{2.363861in}{1.451556in}}{\pgfqpoint{2.363861in}{1.484889in}}%
\pgfpathlineto{\pgfqpoint{2.363861in}{1.932889in}}%
\pgfpathquadraticcurveto{\pgfqpoint{2.363861in}{1.966222in}}{\pgfqpoint{2.330528in}{1.966222in}}%
\pgfpathlineto{\pgfqpoint{0.918194in}{1.966222in}}%
\pgfpathquadraticcurveto{\pgfqpoint{0.884861in}{1.966222in}}{\pgfqpoint{0.884861in}{1.932889in}}%
\pgfpathlineto{\pgfqpoint{0.884861in}{1.484889in}}%
\pgfpathquadraticcurveto{\pgfqpoint{0.884861in}{1.451556in}}{\pgfqpoint{0.918194in}{1.451556in}}%
\pgfpathclose%
\pgfusepath{stroke,fill}%
\end{pgfscope}%
\begin{pgfscope}%
\pgfsetbuttcap%
\pgfsetroundjoin%
\definecolor{currentfill}{rgb}{0.000000,0.000000,1.000000}%
\pgfsetfillcolor{currentfill}%
\pgfsetlinewidth{1.003750pt}%
\definecolor{currentstroke}{rgb}{0.000000,0.000000,1.000000}%
\pgfsetstrokecolor{currentstroke}%
\pgfsetdash{}{0pt}%
\pgfsys@defobject{currentmarker}{\pgfqpoint{-0.034722in}{-0.034722in}}{\pgfqpoint{0.034722in}{0.034722in}}{%
\pgfpathmoveto{\pgfqpoint{0.000000in}{-0.034722in}}%
\pgfpathcurveto{\pgfqpoint{0.009208in}{-0.034722in}}{\pgfqpoint{0.018041in}{-0.031064in}}{\pgfqpoint{0.024552in}{-0.024552in}}%
\pgfpathcurveto{\pgfqpoint{0.031064in}{-0.018041in}}{\pgfqpoint{0.034722in}{-0.009208in}}{\pgfqpoint{0.034722in}{0.000000in}}%
\pgfpathcurveto{\pgfqpoint{0.034722in}{0.009208in}}{\pgfqpoint{0.031064in}{0.018041in}}{\pgfqpoint{0.024552in}{0.024552in}}%
\pgfpathcurveto{\pgfqpoint{0.018041in}{0.031064in}}{\pgfqpoint{0.009208in}{0.034722in}}{\pgfqpoint{0.000000in}{0.034722in}}%
\pgfpathcurveto{\pgfqpoint{-0.009208in}{0.034722in}}{\pgfqpoint{-0.018041in}{0.031064in}}{\pgfqpoint{-0.024552in}{0.024552in}}%
\pgfpathcurveto{\pgfqpoint{-0.031064in}{0.018041in}}{\pgfqpoint{-0.034722in}{0.009208in}}{\pgfqpoint{-0.034722in}{0.000000in}}%
\pgfpathcurveto{\pgfqpoint{-0.034722in}{-0.009208in}}{\pgfqpoint{-0.031064in}{-0.018041in}}{\pgfqpoint{-0.024552in}{-0.024552in}}%
\pgfpathcurveto{\pgfqpoint{-0.018041in}{-0.031064in}}{\pgfqpoint{-0.009208in}{-0.034722in}}{\pgfqpoint{0.000000in}{-0.034722in}}%
\pgfpathclose%
\pgfusepath{stroke,fill}%
}%
\begin{pgfscope}%
\pgfsys@transformshift{1.118194in}{1.841222in}%
\pgfsys@useobject{currentmarker}{}%
\end{pgfscope}%
\end{pgfscope}%
\begin{pgfscope}%
\pgftext[x=1.418194in,y=1.782889in,left,base]{\sffamily\fontsize{14.000000}{14.400000}\selectfont B-UP=1}%
\end{pgfscope}%
\begin{pgfscope}%
\pgfsetbuttcap%
\pgfsetroundjoin%
\definecolor{currentfill}{rgb}{0.000000,0.000000,1.000000}%
\pgfsetfillcolor{currentfill}%
\pgfsetlinewidth{1.003750pt}%
\definecolor{currentstroke}{rgb}{0.000000,0.000000,1.000000}%
\pgfsetstrokecolor{currentstroke}%
\pgfsetdash{}{0pt}%
\pgfsys@defobject{currentmarker}{\pgfqpoint{-0.034722in}{-0.034722in}}{\pgfqpoint{0.034722in}{0.034722in}}{%
\pgfpathmoveto{\pgfqpoint{-0.034722in}{-0.034722in}}%
\pgfpathlineto{\pgfqpoint{0.034722in}{0.034722in}}%
\pgfpathmoveto{\pgfqpoint{-0.034722in}{0.034722in}}%
\pgfpathlineto{\pgfqpoint{0.034722in}{-0.034722in}}%
\pgfusepath{stroke,fill}%
}%
\begin{pgfscope}%
\pgfsys@transformshift{1.118194in}{1.608889in}%
\pgfsys@useobject{currentmarker}{}%
\end{pgfscope}%
\end{pgfscope}%
\begin{pgfscope}%
\pgftext[x=1.418194in,y=1.550556in,left,base]{\sffamily\fontsize{14.000000}{14.400000}\selectfont B-UP=5}%
\end{pgfscope}%
\end{pgfpicture}%
\makeatother%
\endgroup%

%% file: 04-framework.tex
\label{sec:details:counterex}
This section reports on the relevant steps to instantiate an efficient implementation of the syntax-guided synthesis framework.
First, available implementations of CE generation are too restricted in the variety of properties they support.  
Moreover, the embedding into a CEGIS-loop changes the focus of the CE generation. 
We motivate and report on a selection of changes.
Finally, when moving from the analysis of a single model to a family of models, the well-foundedness criteria need reviewing, which we exemplify for three particularly important criteria. 

\subsection{Better CEs for CEGIS}
\label{sec:algorithmic:extendedmaxsat}
Before diving into any changes, we recap the technique of~\cite{DBLP:conf/atva/DehnertJWAK14} for computing traditional program-level CEs:
\begin{definition}[High-level CEs~\cite{DBLP:conf/atva/DehnertJWAK14}]
  Given a program $\program{P}{} = (\Vars, C)$ and property $\varphi = \mathbb{P}_{\leq \lambda} [\lozenge G]$ s.t.\ $\program{P}{} \not\models \varphi$,  $E \subseteq C$ is a \emph{high-level CE} 
  if $\fixdeadlock({\program{P}{}}_{|E}) \not\models \varphi$.
\end{definition}
We have the following connection between high-level and program-level CEs.
\begin{proposition}
  \label{prop:counterex_traditional_iff_generalisable}
  For safety properties, high-level and  program-level CEs coincide.
\end{proposition}
Each program-level CE is trivially a high-level CE. 
A high-level CE is a program-level CE, 
as any additional command can only be enabled in unreachable or deadlock states.
Otherwise, the program would contain overlapping commands, violating Def.~\ref{def:genCE}.
Recall that assuming only non-overlapping programs is crucial.
Consider adding the command \mbox{\texttt{s=0 -> s'=2}} to the program in Fig.~\ref{fig:runningex:counterex}. 
The probability to reach \texttt{s=3} is now \emph{reduced} to $0.25$ resulting in the extended program no longer violating~$\Phi$. Unreachable states are irrelevant to any property and, intuitively speaking, adding transitions to deadlocks cannot decrease the probability to reach $\psi$.

The technique in \cite{DBLP:conf/atva/DehnertJWAK14} computes \emph{minimal} high-level CEs (i.e. the smallest set of commands) violating a given reachability property with an upper bound $\lambda$.
The set is not unique in general.
It reduces the computation to repeatedly solving \texttt{MaxSat} instances over two sets of propositional formulae, $\Xi$ and $\Upsilon$.  
$\Xi$ encodes a selection of commands and (via a negation) the \texttt{MaxSat} solver minimises the size of the command set.  
$\Upsilon$ encodes required (but not sufficient) constraints on valid CEs, e.g, it states that a command enabled in the initial state must be selected. 
In this way, the \texttt{MaxSat} solver returns a candidate set $E$ of commands.
Using standard procedures it is then checked whether
$E$ is sufficient to exceed
$\lambda$. 
If so, $E$ by construction is a minimal CE. 
Otherwise, $\Upsilon$ is strengthened to exclude $E$ and (possibly) other candidate sets.

\subsubsection{More properties}
We provide support for program-level CEs beyond upper bounds on reachability probabilities.
We deem support for liveness properties essential.
Consider a sketch for a controller that moves a robot.
A sketch that includes the option to wait together with the single objective that something bad may only happen with a small probability might simply never make any progress.
While not doing anything typically induces safety, the synthesised controller is not useful.
Likewise, performance criteria such as expected time to completion or expected energy consumption are widespread, and typically expressed as expected rewards.
\paragraph*{Liveness properites}
Proposition~\ref{prop:counterex_traditional_iff_generalisable} is crucial for the correctness of our approach, but it does not hold for lower bounds (e.g. $\varphi = \mathbb{P}s_{\geq \lambda}[\lozenge G]$).
Suppose $E$ is a (traditional) CE for $\program{P}{}$ and $\varphi$. 
Adding commands may add probability-mass that reaches the target and thus may yield a program $\program{P'}{}$ such that $\program{P'}{} \models \Phi$.
However, the statement does hold for lower bounds if there are no deadlocks reachable~in~$\semantics{{\program{P}{}}_{|E}}$.

The idea thus is to trap more probability mass than $1-\lambda$ in states $B$ from which there is no path to the target states $\psi$.
To this end, we first compute a traditional CE $E$ for reaching $B$ with probability at most $1 - \lambda$.
Then, we extend $E$ with commands that ensure that the states in $B$ actually may never reach $\psi$.
This is always possible by construction of $B$.
The resulting set of commands is a program-level CE as in Def.~\ref{def:genCE}.
These CEs are no longer necessarily the smallest (in terms of the number of commands) for given $\program{P}{}$ and $\varphi$.

\begin{example}
Consider the property $\varphi = \mathbb{P}_{>0.6} [\lozenge~\texttt{s=2}]$. The property is violated by the program from  Fig.~\ref{fig:runningexample:instance}.
Fig.~\ref{fig:lowerboundcounterex} constitutes a CE since with probability $0.5$ the system moves to $\texttt{s=3}$ that in turn is equipped with a self-loop.
Hence, the probability to ever reach \texttt{s=2} is guaranteed to be below $0.5$.
Dropping the command with guard \texttt{s=3} is insufficient for this guarantee: an additional command could add transitions from \texttt{s=3} and \texttt{s=1} to \texttt{s=2} and meet the (lower) bound of $\varphi$.
\end{example}

\paragraph*{Expected rewards}
We adapted the ideas of state-level CEs for rewards from~\cite{DBLP:conf/fm/QuatmannJDWAKB15} to the program level. 
Intuitively, in the underlying MC of the program, we replace the self-loops from the $\fixdeadlock$ operation by transitions to a target state.
For its soundness, the method has to assume that for each realisation, the probability to reach the target is one. Otherwise, the costs are by definition infinite, and the instance can be rejected.
Support for lower bounds on rewards requires additional assumptions and has not been considered so far.

\subsubsection{A changed minimisation objective.}
The following adaptions allow us to derive conflicts that are in many situations smaller than the conflicts derived from the minimal CEs described above. Smaller conflicts lead to more effective pruning and thus improve the performance of the framework.
The conflicts we obtain are always at least as good as before.
\paragraph*{Relevant commands}
The \texttt{MaxSat} approach in~\cite{DBLP:conf/atva/DehnertJWAK14} minimises over \emph{all} commands.
Recall that the design space is given only by \emph{relevant commands}, i.e.~the commands containing holes. Therefore, we change $\Xi$ to minimise the number of relevant commands.
All other commands can be added to a CE without negatively affecting the sizes of the generated conflicts.
This significantly reduces the number of candidates considered during the CE generation.
Formula
$\Upsilon$ still reasons about all commands, which allows to restrict the candidate command sets
further.

\begin{example}
In Fig.~\ref{fig:runningexample:sketch} only the commands with guards \texttt{s=0} and \texttt{s=1} are relevant.
Instead of considering up to $2^4$ candidate command sets (all four commands), we only consider up to $2^2$ CEs (the two relevant commands).

\end{example}

\paragraph*{Multiple CEs}
CEs (even with minimal size) are not unique.
For the same realisation, multiple CEs may lead to different conflicts.
Each of them can be used to prune future realisations from being explored.
Furthermore, 
our approach does not require to use only the smallest CE, we may want to explore other CEs beyond the first one.
However, minimality is crucial: If a strict subset of a CE $E$ is already a CE, conflicts generated by $E$ do not prune the design space further.
We thus extend the \texttt{MaxSat} loop to prevents all supersets of $E$ being checked:
If we find a CE $E$, we extend $\Upsilon$ by blocking all CEs which include the set of relevant commands in $E$.

\paragraph*{CEs to multiple properties}
We search for CEs that violate any of the properties.
Candidate CEs that subsume other CEs (potentially for other properties) are superfluous and should not be considered.
We integrated support for CE generation for multiple properties with the same path formula. 
For other combinations, a naive combination induces a severe performance hit to the \texttt{MaxSat} solver, and proper support is left as future work.

\subsection{Handling ill-formed realisations.}
Model checkers typically make some assumptions on the well-foundedness of the input, and might or might not enforce them.
Relevant examples for non well-founded models are the existence of deadlocks in the model, variables that go out of their bounds, and --in this context-- guards that overlap.
In the presence of a sketch, these well-foundedness criteria are easily violated and thus rigorous handling is essential.
 We argue that it is not good to simply throw an error if the well-foundedness is violated by any realisation:
 Such rigour makes sketching overly complicated, and might require manually adding various constraints. Conversely, checking the well-foundedness of every candidate realisation on side of the verifier can significantly degrade the performance as by default only single realisations can be removed from the design space.
 Below we review how we handle ill-formed realisations to achieve a good trade-off between the usability and performance of the synthesis framework.

\paragraph*{Out-of-bounds}
Variable bounds should be strictly respected.
However, constraining the sketch to prevent generation of realisations with out-of-bounds violations might be very complex. 
Therefore, we extend the model checker to add a state which reflects a variable being out-of-bounds by checking the variable values during the MC construction.
Then, the properties are extended with a statement that no probability mass should reach these states. 
This extension allows us to apply conflict analysis and typically prune a set of realisations violating the variable bounds using a single verification query.

\paragraph*{Overlapping guards}
As already pointed out, overlapping guards can prevent effective CE generalisation.
Therefore, we reject realisations with overlapping guards.
In order mitigate the performance degradation caused by returning trivial conflicts, we implement a similar extension as for out-of-bounds.

\paragraph*{Deadlocks}
The existence of deadlocks prevents verification of time unbounded properties, therefore most verification tools automatically apply a $\fixdeadlock$ operation. 
While this operation is sensible for a single program, it is not for sketches with holes in guards, as
deadlocks are problematic for CE generalisation in combination with lower-bounded properties.
Thus, we treat them analogous to out-of-bounds and overlapping guards.